\documentclass[11pt,letterpaper]{article}
\input setup/preamble
\usepackage{amsthm}

\newtheorem{counter}{Counter}[section]
\newtheorem{theorem}[counter]{Theorem}

\newtheorem{lemma}[counter]{Lemma}
\newtheorem{proposition}[counter]{Proposition}
\newtheorem{claim}[counter]{Claim}
\newtheorem{fact}[counter]{Fact}

\newtheorem{definition}[counter]{Definition}

\newtheorem{remark}[counter]{Remark}

\let\emptyset\varnothing

\newcommand\restr[2]{{
  \left.\kern-\nulldelimiterspace 
  #1 
  \vphantom{\big|} 
  \right|_{#2} 
  }}

\DeclareMathOperator{\prob}{\mathbf{P}}
\DeclareMathOperator{\expt}{\mathbb{E}}

\newcommand{\patharrow}{,}

\newcommand{\ppoly}{\mathrm{poly}}
 
\newclass{\OPP}{OPP}
\newclass{\OP}{OP}
\newclass{\BPEXP}{BPEXP}
\newclass{\coNTIME}{coNTIME}

\newlang{\sat}{Sat}
\newlang{\OV}{OV}
\newlang{\LCS}{LCS}
\newlang{\HS}{HS}
\newlang{\maxtwosat}{Max-2-Sat}
\newlang{\maxcut}{MAXCUT}
\newlang{\perm}{Permanent}
\newlang{\minimumHammingDistance}{Minimum Hamming Distance}
\newlang{\formulaSat}{Formula Satisfiability}
\newlang{\satisfiability}{Satisfiability}
\newlang{\uniquesat}{unique-sat}
\newlang{\csp}{Csp}
\newlang{\maxsat}{MaxSat}
\newlang{\maxflow}{MaxFlow}
\newlang{\maximumMatching}{Maximum Matching}
\newlang{\cnfsat}{Cnf-Sat}
\newlang{\cktsat}{Circuit Sat}
\newlang{\turingsat}{Turing Sat}
\newlang{\ind}{Independent Set}
\newlang{\maxind}{Max Independent Set}
\newlang{\subiso}{Subgraph Isomorphism}
\newlang{\hamp}{Hamiltonian Path}
\newlang{\factoring}{Factoring}
\newlang{\graphIsomorphism}{Graph Isomorphism}
\newlang{\primality}{Primality}
\newlang{\parityfunction}{Parity}
\newlang{\hamc}{Hamiltonian Cycle}
\newlang{\clique}{Clique}
\newlang{\colorability}{Colorability}
\newlang{\setsplitting}{Set Splitting}
\newlang{\hittingSet}{Hitting Set}
\newlang{\vertexcover}{Vertex Cover}
\newlang{\independentset}{Independent Set}
\newlang{\feedbackvertexset}{Feedback Vertex Set}
\newlang{\longestpath}{Longest Path}
\newlang{\dominatingset}{Dominating Set}
\newlang{\listcoloring}{List Coloring}
\newlang{\chromaticnumber}{Chromatic Number}
\newlang{\setcover}{Set Cover}
\newlang{\lcs}{Longest Common Subsequence}

\newlang{\subsetChain}{Maximum Length Chain of Subsets}
\newlang{\apsp}{All-Pairs Shortest Paths}
\newlang{\editDistance}{Edit Distance}
\newlang{\orthogonalVectors}{Orthogonal Vectors}
\newlang{\vectordomination}{Vector Domination}
\newlang{\frechetdistance}{Fr\'{e}chet Distance}

\newcommand{\trn}{\tau}
\newcommand{\ntrn}{\#\Gamma}
\newcommand{\trans}{\Gamma}

\newcommand{\total}{\psi}

\renewcommand{\clique}{{\mathbb K}}

\newcommand{\cC}{\mathcal{C}}
\newcommand\enumballsat[2]{\textsc{Enum($#1$, $#2$)}}

\newcommand{\ts}{\textsc{TreeSearch}}

\newcommand{\eqdef}{\stackrel{\textrm{def}}{=}}

\newcommand\enumballsatnae[2]{\textsc{NAE-Enum($#1$, $#2$)}}

\DeclareMathOperator{\maj}{\mathsf{Maj}}

\newcommand{\stage}{\kappa}

\usepackage{enumitem}
\newlength\caselen
\newlist{casesenum}{enumerate}{2}
\setlist[casesenum,1]{label=\textbf{Case~\arabic*.}, 
  itemindent=*,leftmargin=0pt}
\setlist[casesenum,2]{label=\textbf{Case~\roman*.}, 
  itemindent=*,leftmargin=\parindent}

\crefname{paragraph}{Paragraph}{Paragraphs}

\title{Local Enumeration: The Not-All-Equal Case} 

\author{Mohit Gurumukhani\thanks{Cornell University, Ithaca, NY, USA. Supported by NSF CAREER Award 2045576 and a Sloan Research Fellowship. Email: \texttt{mgurumuk@cs.cornell.edu}}
\and
Ramamohan Paturi\thanks{Department of Computer Science and Engineering, University of California, San Diego. Partially supported by NSF grant 2212136. Email: \texttt{rpaturi@ucsd.edu}}
\and
Michael Saks\thanks{Department of Mathematics, Rutgers University, Piscataway, NJ, USA. Email: \texttt{saks@math.rutgers.edu}}
\and
Navid Talebanfard\thanks{University of Sheffield, Sheffield, UK. Email: \texttt{n.talebanfard@sheffield.ac.uk}}
}

\begin{document}
\maketitle

\begin{abstract}

Gurumukhani et al. (CCC'24) proposed the \emph{local enumeration} problem \enumballsat{k}{t} as an approach to break the Super Strong Exponential Time Hypothesis (SSETH): for a natural number $k$ and a parameter $t$, given an $n$-variate $k$-CNF with no satisfying assignment of Hamming weight less than $t(n)$, enumerate all satisfying assignments of Hamming weight exactly $t(n)$. Furthermore, they gave a randomized algorithm for \enumballsat{k}{t} and employed new ideas to analyze the first non-trivial case, namely $k = 3$. In particular, they solved \enumballsat{3}{\frac{n}{2}} in expected $1.598^n$ time. A simple construction shows a lower bound of $6^{\frac{n}{4}} \approx 1.565^n$.

In this paper, we show that to break SSETH, it is sufficient to consider a \emph{simpler} local enumeration problem \enumballsatnae{k}{t}: for a natural number $k$ and a parameter $t$, given an $n$-variate $k$-CNF with no satisfying assignment of Hamming weight less than $t(n)$, enumerate all Not-All-Equal (NAE) solutions of Hamming weight exactly $t(n)$, i.e., those that satisfy and falsify some literal in every clause. We refine the algorithm of Gurumukhani et al. and show that it \emph{optimally} solves \enumballsatnae{3}{\frac{n}{2}}, namely, in expected time $\poly(n) \cdot 6^{\frac{n}{4}}$.

\end{abstract}

\section{Introduction}

Four decades of research on the exact complexity of $k$-SAT has given rise to a handful of non-trivial exponential time algorithms, i.e., algorithms running in time $2^{(1 - \epsilon)n}$ with non-trivial \emph{savings} $\epsilon > 0$ \cite{MonienS85,PaturiPZ99,DSchoning02,DantsinGHKKPRS02,PaturiPSZ05,Hertli14,DBLP:conf/stoc/HansenKZZ19,Scheder21}. Despite extensive effort, PPSZ \cite{PaturiPSZ05} remains essentially the fastest known $k$-SAT algorithm. It is also known that its analysis cannot be substantially improved \cite{DBLP:conf/coco/SchederT20}. The \emph{Super Strong Exponential Time Hypothesis (SSETH)} formalizes the lack of progress in improving the exact complexity of $k$-SAT, and states that it cannot be solved in time $2^{(1 - \omega(1/k))n}$  \cite{VyasW21}. Gurumukhani et al. \cite{GurumukhaniPaturiPudlakSaksTalebanfard_2024_CCC} recently proposed the \emph{local enumeration} problem as a new approach to refute SSETH. More precisely, $\enumballsat{k}{t}$ is defined as follows: for natural number $k$ and a parameter $t$, given an $n$-variate $k$-CNF $F$ with no satisfying assignment of Hamming weight less than $t(n)$, enumerate all satisfying assignments of Hamming weight exactly $t(n)$. They provided a randomized branching algorithm and presented novel ideas to analyze it for the first non-trivial case, $k = 3$. 

In this paper, we argue that fast algorithms for an easier local enumeration problem would also refute SSETH. To be precise, we consider the \enumballsatnae{k}{t} problem which is formally defined as follows: for a natural number $k$ and a parameter $t$, given an $n$-variate $k$-CNF $F$ with no satisfying assignment of Hamming weight less than $t(n)$, enumerate all Not-All-Equal (NAE) solutions of Hamming weight exactly $t(n)$ \footnote{we emphasize that we require $F$ to have no satisfying assignment, not only no NAE-satisfying assignment of weight less than $t(n)$}. Recall that a NAE solution to a CNF is one which satisfies and falsifies a literal in every clause. We will show that a refinement of the algorithm of \cite{GurumukhaniPaturiPudlakSaksTalebanfard_2024_CCC} can \emph{optimally} solve \enumballsatnae{k}{\frac{n}{2}} for $k=3$. For this algorithm $t = \frac{n}{2}$ is the hardest case, and we therefore have an algorithm for all $t \le \frac{n}{2}$. Our proof utilizes a new technique for analyzing a transversal tree (a tree of satisfying solutions with Hamming weight exactly $t(n)$ constructed using the clauses of the $k$-CNF). 

It is easy to see that $k$-SAT can be reduced to $(k+1)$-NAE-SAT by the following folklore reduction: Given a $k$-CNF $F$, define a $(k+1)$-CNF $F'$ as follows. Let $z$ be a new variable. For each clause $C \in F$ we include the clause $C \vee z$ in $F'$. It is clear that $F$ is satisfiable iff $F'$ has a NAE solution\footnote{A satisfying assignment of $F$ can be extended to a NAE solution of $F'$ by setting $z = 0$. Conversely, a NAE solution of $F'$ is a satisfying assignment of $F$ projected on the first $n$ variables if $z = 0$, and if $z = 1$, the negation of the remaining variables is a satisfying assignment of $F$.}. Furthermore, if we can solve $k$-NAE-SAT in time $2^{(1 - \mu_k)n}$, then we can solve $k$-SAT in time $O(2^{(1 - \mu_{k - 1})n})$.
In the other direction, the $k$-NAE-SAT problem has a trivial reduction to $k$-SAT: Given a $k$-CNF $F$, construct the formula
$F'$ by adding, for each clause of $F$ the clause consisting of the negations of its literals, then $F$ has an NAE solution if and only if $F'$ is satisfiable. Therefore, for large $k$, $k$-SAT can be solved with asymptotically the same \emph{savings} as that of $k$-NAE-SAT.


It is noted in \cite{GurumukhaniPaturiPudlakSaksTalebanfard_2024_CCC} that upper bounds on \enumballsat{k}{t} for all $t \le \frac{n}{2}$ imply $k$-SAT upper bounds. It is easy to see that the same holds for \enumballsatnae{k}{t} and $k$-NAE-SAT. Therefore, by the discussion above, as far as $k$-SAT savings are concerned, we may only focus on \enumballsatnae{k}{t} for all $t \le \frac{n}{2}$.

\paragraph{Lower bounds for local enumeration} Define $k$-CNF $\maj_{n, k}$ as follows. Divide the $n$ variables into blocks of size $2k - 2$, and for each block include all positive clauses of size $k$. Every satisfying assignment of $\maj_{n, k}$ sets at least $k - 1$ variables in each block to 1, and thus has Hamming weight at least $\frac{n}{2}$. Notably, all minimum-weight satisfying assignments are NAE. Furthermore, the number of minimum-weight satisfying assignments of $\maj_{n, k}$ is $2^{(1 - O(\log(k)/k))n}$. This, in particular, is a lower bound on the complexity of both \enumballsat{k}{\frac{n}{2}} and \enumballsatnae{k}{\frac{n}{2}}, and thus if we can show that any of these bounds is tight for all $t \le \frac{n}{2}$, we break SSETH. We reiterate that it is wide open to obtain even a combinatorial (non-algorithmic) upper bound of this kind; while this is not good enough to break SSETH, it is necessary and will be very interesting to obtain such a bound.

\subsection{Our Contributions}

Gurumukhani et al. \cite{GurumukhaniPaturiPudlakSaksTalebanfard_2024_CCC} showed that their algorithm solves \enumballsat{k}{t} for all $t \le \frac{n}{2}$, and in particular, it solves \enumballsat{3}{\frac{n}{2}} in expected $1.598^n$ time. Compare this with the lower bound of $6^{\frac{n}{4}} \approx 1.565^n$ which follows from $\maj_{n, 3}$. With respect to this algorithm, the complexity of \enumballsat{k}{t} increases as $t$ grows. Therefore $t = \frac{n}{2}$ is the hardest instance for this algorithm. Here we refine their algorithm, and show that it optimally solves \enumballsatnae{3}{\frac{n}{2}}, which is also the hardest instance for this algorithm.

\begin{theorem}
\label{thm:monotone}
For $n\ge 0, t\le n/2$, let $F$ be an arbitrary $n$-variate $3$-CNF where every satisfying assignment has Hamming weight at least $t$. Then, the number of NAE satisfying assignments of $F$ of Hamming weight exactly $t$ is at most $6^{\frac{n}{4}}$. Furthermore, we can enumerate these solutions in expected $\text{poly}(n)\cdot 6^{\frac{n}{4}}$ time.
\end{theorem}

\cref{thm:monotone} will follow from \cref{thm: bound for large t0} and \cref{thm: bound for small t0}.
The algorithm of \cite{GurumukhaniPaturiPudlakSaksTalebanfard_2024_CCC} is a randomized variant of the seminal method of Monien and Speckenmeyer \cite{MonienS85}: take a positive clause $C = x_1 \vee \ldots \vee x_k$ and for a random ordering $\pi$ of its variables, recursively solve the problem under the restriction $x_{\pi(1)} = \ldots = x_{\pi(i - 1)} = 0, x_{\pi(i)} = 1$, for every $i \in [k]$. While the analysis in \cite{GurumukhaniPaturiPudlakSaksTalebanfard_2024_CCC} is essentially local (only depends on the information along root-leaf paths), we had to develop a global technique to get the optimal bound where we use information from one subtree to analyze the performance in an entirely different subtree. We believe that the elements of our analysis have the potential to generalize for $k>3$.

\paragraph{Depth-3 complexity of Majority function.} Majority is a natural candidate function for beating the state-of-the-art depth-3 circuit lower bounds \cite{HastadJP95,LecomteRT22,Amano23Majority}. The local enumeration paradigm of \cite{GurumukhaniPaturiPudlakSaksTalebanfard_2024_CCC} gives a promising paradigm to establish this, and their result yields new lower bounds for $\Sigma^3_3$ circuits, i.e., OR-AND-OR circuits with bottom fan-in at most 3. Our result can also be interpreted as an optimal lower bound Majority for $\Sigma^3_3$ circuits in which every Hamming weigh $\frac{n}{2}$ input is a NAE solution of some depth-2 subcircuit.

\paragraph{Hypergraph Tur{\'a}n problems.} Following \cite{FranklGT22,GurumukhaniPaturiPudlakSaksTalebanfard_2024_CCC, GKP24}, by restricting ourselves to monotone formulas, our result has a natural interpretation for hypergraphs. Let $H = (V, E)$ be a 3-uniform $n$-vertex hypergraph with no transversal of size less than $\frac{n}{2}$, i.e., any set of vertices that intersects every hyperedge has size at least $\frac{n}{2}$. We show that the number of 2-colorings of $H$ with no monochromatic hyperedge is at most $6^{\frac{n}{4}}$ and give a randomized algorithm that in expected $\text{poly}(n) \cdot 6^{\frac{n}{4}}$ time enumerates them. This bound is tight, by considering the disjoint union of $\frac{n}{4}$ cliques of size 4 (this hypergraph corresponds to the formula $\maj_{n, 3}$).

\paragraph{Combinatorial interpretation.} Our result can also be interpreted in a purely combinatorial manner. For instance, let $S(n, t, k)$ be the maximum number of weight $t$ satisfying assignments of a $k$-CNF that has no solutions of weight less than $t$. By construction, we can show $S(n, n/2, 3)\ge 6^{n/4}$. It was conjectured in \cite{GurumukhaniPaturiPudlakSaksTalebanfard_2024_CCC} that $S(n, n/2, 3)\le 6^{n/4}$. We here prove $S(n, n/2, 3)\le 6^{n/4}$ under an additional assumption that the $3$-CNF formula $F$ is negation closed: If $\alpha$ satisfies $F$, then negation of $\alpha$ also satisfies $F$.
As pointed out in \cite{GurumukhaniPaturiPudlakSaksTalebanfard_2024_CCC, GKP24}, proving strong upper bounds for $S(n, n/2, k)$ for $k > 3$ is a major open problem and doing so for large $k$ would lead to breakthrough circuit lower bounds for depth $3$ circuits.
The current best bounds were first obtained by \cite{HastadJP95} who showed $S(n, n/2, k) \le 2^{n - O(n/k)}$.
We also note that to refute SSETH, one not only needs to show strong upper bounds for $S(n, n/2, k)$ but also needs an enumeration algorithm.

\subsection{Proof Strategy}

As mentioned earlier, our enumeration algorithm is similar to \cite{GurumukhaniPaturiPudlakSaksTalebanfard_2024_CCC} where we select an unsatisfied monotone clause $C$, randomly order the variables in $C$ and for $1\le i\le 3$, set the first $i-1$ variables to $0$, set variable $i$ to $1$ and recurse. This gives rise to a recursion tree (henceforth called transversal tree) where each leaf may correspond to a transversal (it could be a leaf corresponds to a falsified formula). We bound the expected number of leaves that our algorithm visits by carefully ensuring we do not double count leaves that correspond to the same transversal. This is also what \cite{GurumukhaniPaturiPudlakSaksTalebanfard_2024_CCC} did. The key difference here is that in our algorithm, we very carefully choose which clause to develop next: we divide the transversal tree into stages and for each stage, carefully argue certain kinds of clauses must exist and carefully pick those clauses to develop the transversal tree. This is also where we use the not-all-equals assumption to show certain kinds of favorable clauses must exist. With careful accounting, we are able to obtain the tight bound.

\section{Transversal trees and the TreeSearch algorithm} \label{sec:preliminaries}
In this section we review the $\ts$ algorithm from \cite{GurumukhaniPaturiPudlakSaksTalebanfard_2024_CCC}
which enumerates the solutions of a $k$-CNF solutions of minimum Hamming weight. We also make some additional observations that allow us to get a tight analysis of the algorithm for NAE-3-SAT.  

The $\ts$ algorithm is for $k$-SAT and we want algorithms for NAE-$k$-SAT. Define the \emph{negation-clause} for clause $C$  to be the clause $C'$ whose literals
are the negations of the literals of $C$. The \emph{negation-closure}
 of a formula $F$ is the formula $\hat{F}$ obtained from $F$ by adding the negation-clause of every clause of $F$ (if it is not already in $F$).  We say that $F$ is negation-closed if $\hat{F}=F$. 
 It is easy to see that the set of NAE solutions for a formula $F$ is exactly the same as the set
 of SAT-solutions for its negation-closure $\hat{F}$.    So  any algorithm that enumerates the minimum-weight
 satisfying assignments
for $k$-CNF formulas can be used to find the minimum weight NAE-assignments of a $k$-CNF formula $F$ by applying the algorithm to $\hat{F}$.

\subsection{Transversals and Transversal trees}

\subsubsection{Important definitions}

\begin{definition}[Transversals]
A set $S\subseteq X$ is a \emph{transversal of $F$} if the assignment that sets
the variables in $S$ to $1$ and the variables in 
$X\setminus S$ to $0$ is a satisfying assignment of $F$.
\end{definition}

We say $S$ is a \emph{minimal transversal of  $F$} if no subset of $S$ is a transversal.
We say that $S$ is a \emph{minimum-size transversal of $F$} if it has the smallest size over all transversals.

The definition of transversal of a formula can be seen as a natural generalization of the standard notion of \emph{transversal
of a hypergraph}, which is a set that has nonempty intersection with every edge. Indeed,
if $F$ is a monotone formula
(where every clause consists of positive literals) then a transversal of $F$ is exactly a transversal
of the hypergraph $H$ consisting of the set of clauses of $F$.

\begin{definition}[Transversal number]
For a satisfiable $k$-CNF $F$, the \emph{transversal number} $\trn(F)$ is the cardinality of the minimum-size transversal of $F$.   The set of all minimum-size
transversals of $F$ is denoted by $\trans(F)$ and the cardinality  of $\trans(F)$ is denoted by 
$\ntrn(F)$.
\end{definition}

We focus on  two related problems: (1) Give an algorithm that enumerates $\trans(F)$ and analyze its complexity, and (2) Determine the maximum cardinality of
$\trans(F)$ among all $k$-CNF (or NAE-$k$-CNF) with $\trn(F)=n/2$.  Our main technical tool will be \emph{transversal trees}, introduced in \cite{GurumukhaniPaturiPudlakSaksTalebanfard_2024_CCC}{}, and defined below.
The definition applies to $k$-CNF with $\trn(F)=t$ for any $k,t$.

We need some preliminary definitions.  We associate a subset $Y$ of variables to the partial
assignment that sets all variables of $Y$ to 1. A clause $C$ is said to be
\emph{satisfied by $Y$} if $C$ contains a positive literal corresponding to a variable from $Y$.

\begin{definition}[Simplification of clause / formula]
For a clause $C$ that is not satisfied by $Y$, the \emph{simplification} of $C$ by $Y$, $C/Y$ is the clause obtained by removing occurrences of negations of  variables of $Y$.
Similarly, the \emph{simplification} of $F$ by $Y$, denoted $F/Y$, is
the formula on $X-Y$ obtained by deleting clauses satisfied by $Y$, and replacing
each remaining clauses $C$ by $C/Y$.
\end{definition}

Note that the underlying set of variables of both $F$ and $F/Y$ is the same - $X$. Furthermore, the set of satisfying assignments of $F$ with variables in $Y$ set to 1 is in 1-1 correspondence with the set of satisfying assignments of $F/Y$. We say that $F$ is \emph{falsified by $Y$} if $F/Y$ contains an empty clause.

We will be considering rooted trees with root $r$ where each edge $e$ is assigned a label $Q(e) \in X$. 
\begin{definition}
For a node $u$ and descendant node $v$ we have the following definitions:
\begin{itemize}
\item  $P(u \patharrow v)$ denotes the path from $u$ to $v$.
\item The \emph{shoot from $u$ to $v$}, $S(u \patharrow v)$, consists of the edges of $P(u \patharrow v)$
together with all child edges of nodes on the path other than $v$.
\item  $Q(u \patharrow v)=\{Q(e):e \in P(u \patharrow v)\}$.  We write $Q(v)$
for $Q(r \patharrow v)$.
\item A clause $C$ is \emph{live at $v$} provided
that $C$ contains no variable in $Q(v)$ (but it may contain the negation of variables in $Q(v)$).  For such a clause we write $C/v$ for $C/Q(v)$
and for a formula $F$, we write $F/v$ for $F/Q(v)$. 
\end{itemize}
\end{definition}

\subsubsection{Constructing the tree}\label{subsubsec: constructing the tree}

We now define a process for growing a tree with node and edge labels $Q(e)$
starting from a root $r$.  The leaves of the current tree are referred to as \emph{frontier nodes}.
Each frontier node $v$ is examined and is either designated as a leaf (of the final tree) or a non-leaf as follows:

\begin{itemize}
\item If $v$ is a node at depth at most $t-1$ such that $F/v$ contains no empty clause, then $v$ is a non-leaf.  It is labeled by a clause $C$ for which $C/v$ is a positive clause\footnote{$F/v$ must contain a positive clause, otherwise $Q(v)$ is a transversal of size less than $t$, contradicting $\trn(F)=t$.}.  Node $v$ is expanded to have $|C/v|$ children with the edges labeled by distinct variables in $C/v$.  We denote the label of an edge $e$ by $Q(e)$.
\item If $F/v$ contains an empty clause then $v$ is a leaf of the final tree called a
     \emph{falsified leaf}.  For the parent $u$ of such a leaf $v$, $F/u$ has no empty clause.  Since $F/v$ is obtained from $F/u$  by setting $Q(uv)$ to 1, the single literal clause $\neg Q(uv)$ is in $F/u$.  We refer  $uv$ as a \emph{falsifying edge}.
\item If $v$ is a node at depth $t$ then $v$ is a leaf.  If it is not a falsified leaf then $v$ is a \emph{viable leaf}.
    
\end{itemize}

A tree constructed according to the above process is a \emph{transversal tree}.  The following easy fact (noted in \cite{GurumukhaniPaturiPudlakSaksTalebanfard_2024_CCC}{}) justifies the name:

\begin{proposition}
    \label{prop:general-tree-existence} 
    Let $F$ be a formula with $\trn(F)=t$ and let $T$ be a transversal tree for $F$.  For every
    minimum-size transversal $S$ of $F$, there is a depth $t$ leaf $v$ of $T$ such that $S=Q(v)$.
\end{proposition}

The depth-$t$ leaves can be divided into three types: falsified leaves,
viable leaves $\ell$ for which $Q(\ell)$ is a transversal,
and viable leaves $\ell$ for which $Q(\ell)$ is not a transversal.  For a leaf of the third type, $Q(\ell)$ might be
a subset of one or more transversals of size larger than $t$.

\begin{remark}
Notice that if a leaf $u$ appears before depth $t$, then $u$ must be a falsified leaf. Unless arising from `effective width $2$ clauses', for our analysis sake, we will continue expanding the tree below $u$. Since $u$ is already falsified, we assume, without loss of generality, that all possible clauses of width at most $3$ are live at $u$ (including the empty clause). We then choose a monotone clause and recursively develop the tree underneath $u$ until we reach depth $t$. Similarly, at all nodes $v$ underneath $u$, we pretend this is the case, that all clauses are live at $v$. We do this since our algorithm will dictate for various nodes which kind of clause must be chosen and developed next. We will argue that certain kinds of clauses exist for such nodes. Hence, to simplify analysis, we will assume all clauses needed are already present.
\end{remark}

Note that it is possible that many leaves correspond to the same transversal.

\subsection{The TreeSearch algorithm for enumerating minimum-size transversals} 

From \cref{prop:general-tree-existence}, we can enumerate all minimum-sized transversals
of $F$ by constructing a transversal tree for $F$, traversing the tree, and for each leaf $\ell$ computing
$Q(\ell)$ and testing whether $Q(\ell)$ is a transversal.
To make this procedure fully algorithmic we need to specify two things. 

The first is a \emph{clause selection strategy} which determines, for each node $v$, the clause of $F$ that labels $v$.
In general, the clause selection strategy may be randomized, but in this paper we consider only deterministic strategies.

The second thing to be specified is the order of traversal of the tree, which is specified  a family $\pi=\{\pi(v):\text{internal nodes $v$ of $T$}\}$ where $\pi(v)$ is a left-to-right
ordering of the variables of $C/v$ and thus of the child edges of $v$.  We use the ordering to determine
a depth first search traversal of the tree which upon arrival at each leaf $\ell$, outputs $Q(\ell)$ if
$Q(\ell)$ is a transversal.  We will say that \emph{$u$ is to the left of $v$} if it is visited before $v$ in the traversal. The running time of the algorithm is dominated by the size of the tree, which
may be as large as $k^{\trn(F)}$.  

To speed up the search, 
\cite{GurumukhaniPaturiPudlakSaksTalebanfard_2024_CCC}{} described a simple criterion on edges such that in the pruned tree obtained by removing
all edges meeting this criterion and the subtrees below, every minimum-size transversal corresponds to a unique leaf,

To formulate this criterion note that
if $u$ is an ancestor of $v$, the edges of
shoot $S(u \patharrow v)$
are situated in one of three ways with respect to
the path $P(u \patharrow v)$: to the left of the tree path, 
to the right of the tree path, or along the tree path.

\begin{definition}[Superfluous edge]
An edge $(uv)$ is \emph{superfluous} if there is an ancestor $w$ of $v$ that has a child edge $wz$ to the left of $P(r \patharrow v)$ such that $Q(wz)=Q(uv)$. 
\end{definition}

We define the algorithm
$\ts$ on the ordered tree $(T,\pi)$ to be the depth first search procedure with the following modification: before
traversing any edge check whether it is superfluous; if it is then
skip that edge (thereby pruning the edge and the subtree below it). This indeed corresponds to setting the corresponding variable to $0$.

\begin{proposition}
\label{prop:exactly once}
(Implicit in \cite{GurumukhaniPaturiPudlakSaksTalebanfard_2024_CCC})
For any transversal tree $T$ of $F$ and any ordering of $T$, the tree  obtained by removing all superfluous
edges and the subtrees below them satisfies: every minimum-size transversal $S$ of $F$ has a unique
leaf $\ell$ of the pruned tree for which $Q(\ell)=S$, and this leaf is the leftmost leaf of the original tree
for which $Q(\ell)=S$. Therefore $\ts$ outputs each minimum-size transversals exactly once.
\end{proposition}

The time complexity of $\ts$ is bounded (up to $\poly(n)$ factors) by the number of leaf nodes in the pruned tree.  The number of leaves depends on the clause selection strategy and the order $\pi$.
Our goal is to choose the strategy and the order so that we can prove a good upper bound on the number of leaves.

In
\cite{GurumukhaniPaturiPudlakSaksTalebanfard_2024_CCC}, 
the algorithm for 3-SAT was analyzed under a specific clause selection rule by considering a randomly chosen ordering $\pi$, in which 
the children of each node in the transversal tree are ordered independently
and uniformly at random.  This enabled them to get an upper bound that is non-trivial, though probably not optimal.  

In this paper, we use a similar approach to analyze NAE-3-SAT. As in \cite{GurumukhaniPaturiPudlakSaksTalebanfard_2024_CCC} we choose $\pi$ at random.
We combine this with a carefully chosen
clause selection strategy, described in 
\cref{section:canonical}, which leads to an optimal upper bound for enumeration of minimum-sized transversals.

\subsection{Pruning under random edge ordering}

Let $F=(X,\cC)$ be a $k$-CNF. 
Let $t=\trn(F)$ be the transversal number of $F$.
Let $T$ be a transversal tree for $F$. As indicated above we consider a random ordering $\pi$ in which
the child edges of each node are randomly ordered from left-right independent of other nodes. Under this distribution, whether or not a particular edge is superfluous is a random event.  
\begin{definition}[Survival and survival probability]
\begin{itemize}
\item[]
\item 
An edge $uv$ is said to \emph{survive} provided that $v$ is not a falsified leaf $v$ and $uv$ is not superfluous.  We say that \emph{path $P(u \patharrow v)$ survives} if every edge of the path survives, and \emph{node $v$ survives} if path $P(r \patharrow v)$ survives.
\item The conditional survival probability of an edge $uv$, $\sigma(uv)$ is defined to be
$\prob[v \text{ survives }|u \text{ survives }]$.  More generally
the \emph{conditional survival probability $\sigma(u,v)$ of path $P(u \patharrow v)$} is defined to be $\prob[P(u\patharrow v) \text{ survives }|u \text{ survives }]$. It follows that $\sigma(u, v)=\prod_{e \in P(u \patharrow v)} \sigma(e)$.
\item For a node $u$, let $L(u)$ be the number of leaves of the subtree $T(u)$ that survive,
and define the \emph{survival value of $u$}, $\total(u)$, to be the conditional expectation $\expt[L(u)|u \text{ survives }]$.  It follows from linearity of expectation
that $\total(u)=\sum_{\text{$v$ is leaf of $T(u)$}} \sigma(u\patharrow v)$.
\end{itemize}
\end{definition}

\begin{proposition}
\label{prop:}
Let $F$ be a 3-CNF with $n$ variables and $\trn(F)=n/2$.  Fix a clause selection rule for building transversal trees.  
    Then the expected running time of $\ts$ on a randomly selected order is $\total(r)\poly(n)$ and the number of transversals of $F$ satisfies $\ntrn(F) \leq \total(r)$.
\end{proposition}
\begin{proof}
For the first statement, the running time of $\ts$ for a given  $\pi$ is
at most $L(r)\ppoly(n)$ and so the expected running time on a random order is at most $\total(r)\ppoly(n)$.
For the second statement, for any order $\pi$, \cref{prop:exactly once} implies that $\ntrn(F) \leq L(r)$, and so $\ntrn(F) \leq \total(r)$.  
\end{proof}

In the sections that follow, we will describe a clause specification strategy and analyze
$\total(r)$ for the case of negation-closed 3-CNF formula (which as noted earlier, 
corresponds to the case of NAE-3-SAT). For the analysis we will need some additional observations.

First, we determine a condition under which we can exactly determine $\sigma(uv)$ for an edge $uv$.
It follows from the definition of survival probability if $uv$ is a falsifying edge then $\sigma(uv)=0$.
So we consider the case that $uv$ is not falsifying.  None of the edges on the path
$P(r \patharrow u)$ are falsified (since the child node of a falsifying edge is always a leaf).

If $uv$ is not falsifying then it survives if and only if is not superfluous.  The condition
for being superfluous depends on the label of $uv$ appearing as the label of a child edge of an ancestor of $u$.
To  account for this we use a system of marking of edges and vertices:

\begin{definition}[Marking of edges and vertices]

\begin{itemize}
\item[]
\item
The \emph{marking set} $M(e)$ of a tree edge $e=(u,v)$  is the set of
nodes $w\neq u$ in the path $P(r\patharrow u)$ which have a child edge $e'$ such that $Q(e)=Q(e')$.
For  $w \in M(e)$ we say that \emph{$w$ marks  the edge $e$} and \emph{$w$ marks the label $Q(e)$}.
\item An edge $e$ is \emph{marked} provided that $M(e) \neq \emptyset$.
\item A node $u$ is \emph{$i$-marked} (for $i \in \{0,1,2,3\}$) if exactly $i$ child edges of $u$ are marked.
\end{itemize}
\end{definition}

Marked edges are edges that have a non-zero probability of being superfluous. The analysis
in \cite{GurumukhaniPaturiPudlakSaksTalebanfard_2024_CCC} uses this to obtain upper bounds on the survival
probability of nodes. We now show that under 
favorable conditions we can calculate the survival probability of nodes exactly.

The event that edge $uv$ survives is exactly the event that for every node $w \in M(uv)$,
 the child edge of $w$ with the same label as $uv$ is to the right of the path edge that comes from $w$, which
 happens with probability 1/2.
 Thus whether $uv$ survives is determined by the orderings $\pi(w)$ of the child edges of $w$ for all $w \in M(uv)$,
 and $\prob[uv \text{ survives }]=2^{-|M(e)|}$.

 The above computation gives the \emph{unconditioned} probability that $uv$ survives, but $\sigma(uv)=\prob[uv \text{ survives }|u \text{ survives }]$ is a conditional probabilty.  We now
 identify a condition under which the conditioning event is independent of the event that $uv$ survives.
Say that an edge $uv$ is \emph{disjointly marked} if its marking set $M(uv)$ is disjoint
from $M(e)$ for every edge $e \in P(r \patharrow u)$.  For a disjointly marked edge, the event that $uv$ survives is independent of the event
that $u$ survives, i.e., that all edges on $P(r \patharrow u)$ survive, 
because the latter event is determined
by the order $\pi(w)$ for nodes that mark some edge of $P(r \patharrow u)$ and since $uv$ is
disjointly marked, this set of nodes is disjoint from $M(uv)$.  Since the order of child edges
of nodes are chosen independently, the event that $uv$ survives is
independent of the event that all edges on $P(r \patharrow u)$ survive.
From this we conclude:

\begin{proposition}
\label{prop:disjointly marked 1}
Let $T$ be a transversal tree for the k-CNF $F$. 
\begin{enumerate}
     \item Let $uv$ be an edge that is not a falsifying edge and is disjointly marked.Then
$\sigma(uv)=2^{-|M(e)|}$.
\item If $v$ is a leaf that is not falsifying and each edge on $P(r \patharrow v)$ is disjointly marked then $\sigma(v)=2^{-\sum_{e \in P(r \patharrow v)} |M(e)|}$. 
\end{enumerate}
\end{proposition}

For the case of negation-closed 3-CNF (which corresponds to NAE 3-SAT) we note the following:

\begin{proposition}
\label{prop:disjointly marked 2}
If $F$ is a  negation-closed 3-CNF formula then in any transversal tree $T$, every edge that
is not a falsifying edge is disjointly marked.
Therefore \cref{prop:disjointly marked 1} applies to every edge and every leaf of $T$.
\end{proposition}

\begin{proof}
Let $uv$ be an edge with ancestor edge $wz$ and $y$ is a node that marks both $uv$ and $wz$. We claim that $uv$ is a falsifying edge. $y$ is necessarily an ancestor of $w$. Let $y'$ be the child of $y$ on the path to $w$. Then the clause labeling $y$ is $Q(yy') \vee Q(wz) \vee Q(uv)$.  Since $F$ is negation closed, $F$ also contains the clause $(\neg Q(yy')) \vee (\neg Q(wz)) \vee (\neg Q(uv))$.  This clause is falsified at the node $v$ since $Q(yy'),Q(wz),Q(uv) \in Q(v)$, and therefore $v$ is a falsifying leaf.
\end{proof}

We need a few additional definitions.

\begin{definition}[Mass of a node]
For a non-leaf node $u$ in a transversal tree,
the \emph{mass of $u$} is the conditional expectation of the number of surviving children of $u$ given that $u$ survives.  By linearity of expectation this is the same as $\sum_{v: \text{ child node of } u} \sigma(u, v)$.  We also refer to this as the \emph{mass of the clause at $u$}.
\end{definition}

\begin{definition}[Defect]
For node $u$, with $e$ child edges, let \emph{defect} of $u$ to be $3 - e$.
We similarly define \emph{defect} of path / shoot as the sum of the defects of all nodes on the path / shoot.
\end{definition}

\begin{definition}[Weight]
Let $S(u\patharrow v)$ be a shoot in $T$. 
The \emph{weight} of $S(u\patharrow v)$ is defined as 
$W(u\patharrow v) \eqdef |\{e \in S: M(e) \ne \emptyset \}|+ \textrm{ defect of } S(u\patharrow v)$,
i.e., the number of marked edges along $S(u\patharrow v)$ plus the defect of $S(u\patharrow v)$. 
\end{definition}

The following fact provides a lower bound on the weight of each root to leaf
shoot in $T$.

\begin{fact}
\label{fact:shoot-extended-weight-lower-bound}
Every root to leaf shoot $S(r\patharrow u)$ in $T$ has a weight of at least $3t - n$.
\end{fact}

\begin{proof}
Let $f$ be the defect of $S(r\patharrow u)$. 
Since the depth of $T$ is $t$,
a root to leaf shoot has $3t - f$ edges. Partition the edges of the shoot into classes according
to the label of the edge, and note that all edges in a class are at different
depths.  For each class, the edge  of minimum depth is unmarked and all other edges in
the class are marked, so the number of unmarked edges is at most $n$ and the number of marked
edges is at least $3t-f-n$. Hence, the weight of $S(r\patharrow u)$ is at least $(3t-f-n) + f = 3t-n$.
\end{proof}


\section{Clause Selection Criterion}
\label{section:canonical}

We here provide a clause selection criterion  for transversal trees for 3-CNFs that will provide us with a tight analysis for the number of NAE-solutions and for the complexity of enumerating them. The clause selection is divided into three stages. Each node will be associated with a stage and a corresponding clause will be selected from that stage. The three stages are:
\begin{enumerate}
    \item \emph{Disjoint stage}
    \item \emph{Controlled stage}
    \item \emph{Arbitrary stage}
\end{enumerate}

We begin the disjoint stage by selecting a pseudomaximum collection $\cC^0$ of disjoint monotone clauses of width $3$ from $F$(See \cref{rem:matching} for what we mean by pseudomaximum collection; on first reading, the readers should pretend as if we selected maximum such collection). Let $t_0 = |\cC^0|$. Our further clause selection criterion depends on the value of $t_0$:
\begin{itemize}
\item 
If $t_0 \ge \frac{n}{4}$, then we skip the controlled stage and directly enter the arbitrary stage where we select arbitrary monotone clauses for the rest of the $\ts$ process.
\item 
If $t_0 \le \frac{n}{4}$, then we enter the controlled stage where we carefully control which clauses to select, argue about existence of certain clauses and our analysis leverages this control. This controlled stage also helps us get a handle on the kinds of clauses we can encounter in the arbitrary stage.
\end{itemize}

We here describe the disjoint stage further. The arbitrary stage needs no further explanation. We analyze the $t_0 \ge \frac{n}{4}$ case in \cref{sec: bound for large t0}. We will explain the controlled stage further in \cref{sec: controlled} and will analyze the $t_0 \le \frac{n}{4}$ case depending upon it in \cref{sec: bound for small t0}.

\subsection{The disjoint stage}\label{subsec: the disjoint stage}

Let the clauses in $\cC^0$ be arbitrarily ordered as $C^0_0, C^0_1, \ldots, C^0_{t_0-1}$.
We will develop the transversal tree so that for all $0\le i\le t_0-1$, all nodes at level $i$ will be expanded using clause $C^0_i$.  We can do this because the disjointness of
the clauses ensures that $C^0_i$ is a live clause at every node at level $i$.
We record the useful observation that any clause used to develop a node appearing at level $\ell \ge t_0$ will have at least one marking:

\begin{fact}\label{fact: after level t0 will always have marking}
Let $u$ be a node at level $\ell \ge t_0$ and let $C$ be a width $3$ clause used to expand at $u$. Then, $u$ will not be $0$-marked, i.e., $u$ will be $j$-marked for $1\le j\le 3$.
\end{fact}
\begin{proof}
For $u$ to be $0$-marked, $C$ must be a width $3$ monotone clause and $C$ must be disjoint from all clauses from $\cC^0$. However, that contradicts the fact that $\cC^0$ is a maximal collection of disjoint width $3$ monotone clauses.
\end{proof}

\begin{remark}[Algorithmic aspect of maximum matching - pseudomaximum matching]\label{remark: maximum size cc0}
\label{rem:matching}
Finding a collection $\cC^0$ of maximum size is NP-hard, and may require large exponential overhead. To mitigate this, we implement an auxiliary data structure that maintains a pseudomaximum collection of disjoint clauses. Initially, the data structure will greedily pick a maximal such collection of clauses. Later in the algorithm, we might encounter scenarios where we can find a larger collection of disjoint clauses. In these cases, we will `reset' the algorithm with that larger collection of disjoint clauses. Such a reset can take place at most $n$ times and hence, the overhead is $\poly(n)$. In our analysis, we will make claims that take this for granted and allow them to use the fact that $\cC^0$ has maximum size ; for any claim where we use this, if the claim does not hold because $\cC^0$ is not necessarily of maximum size, the proof of the claim actually will supply us with clauses that will help us form a larger disjoint collection of clauses and we will `reset' our algorithm.
\end{remark}


\section{Bounding \texorpdfstring{$\total$}{psi} when \texorpdfstring{$t_0 \geq  n/4$}{t0 >= n/4}}\label{sec: bound for large t0}\label{sec: large t0}

In this section, we will show that if in a transversal tree, the number of maximally disjoint clauses chosen in the disjoint stage is at least $n/4$, then $\total(r) \le 6^{n/4}$ where $r$ is the root of the tree. Formally:

\begin{theorem}\label{thm: bound for large t0}
Let $T$ be a transversal tree developed using the clause selection criterion as laid out in \cref{section:canonical} with $t_0(T) \ge \frac{n}{4}$.
Then, $\total(r)\le 6^{\frac{n}{4}}$ where $r$ is the root of $T$.
\end{theorem}


We will use the following lemma to prove the theorem:

\begin{lemma} \label{lem: bounding mwd with fwd}
Let $0\le d\le \frac{n}{2} - t_0, w\ge 0$ and let $u$ be a node at level $\frac{n}{2} - d$ such that every $u$ to leaf shoot in $T(u)$ has weight at least $w$.
Then, $\total(u)\le F(w, d)$ where $F(w, d)$ is defined as follows:

\begin{align*}
F(w,d) &=
\begin{cases}
(\frac{5}{2})^{2d-w} 2^{w-d}& \text{ if $w\le 2d$}\\
2^{3d-w} (\frac{3}{2})^{w-2d} & \text{ if $2d\le w$}.\\
\end{cases}
\end{align*}
\end{lemma}

Assuming this lemma, our main theorem easily follows:

\begin{proof}[Proof of \cref{thm: bound for large t0}]
Let $t_0 = \frac{n}{4} + \Delta$ for some $\Delta \ge 0$.
Let $u$ be an arbitrary node at depth $t_0$.
Let $T(u)$ be the sub-tree rooted at $u$.
The tree $T(u)$ has remaining depth $\frac{n}{4} - \Delta$, and minimum weight of every root to leaf shoot is at least $\frac{n}{2}$ since all edges in the disjoint stage are unmarked.
Thus, $T(u)$ satisfies the requirements of \cref{lem: bounding mwd with fwd}, and we infer that $\total(u)\le F(n/2, n/4 - \Delta)$.
Summing over all $3^{n/4 + \Delta}$ nodes at the end of the disjoint stage, we infer that
\begin{align*}
\total(r)
& \le 3^{n/4+\Delta} F(n/2, n/4 - \Delta)\\
& = 3^{n/4+\Delta} 2^{n/4 - 3\Delta}\left(\frac{3}{2}\right)^{2\Delta}\\
& = 6^{n/4}\left(\frac{27}{32}\right)^{\Delta}\\
& \le 6^{n/4}
\end{align*}
where the last inequality follows because $\Delta \ge 0$.
\end{proof}

\section{Clause Selection Criterion: the Controlled Stage}
\label{sec: controlled}

Recall that this stage appears after the disjoint stage and applies to nodes $u$ at level $t_0$ or larger provided $t_0 \le \frac{n}{4}$. 
To analyze this stage we will need to construct the transversal tree more carefully, using an involved clause selection criterion which is described in this section.  

We use the structural properties of the tree to obtain our lower bound in \cref{sec: bound for small t0}.

In the controlled stage, we will carefully select clauses in a way that imposes useful restrictions on the structure of the tree. After that, we enter the arbitrary stage during which we expand the tree using monotone clauses in an arbitrary order. We will show that the structure of the clauses in the controlled stage imposes useful restrictions on the structure of the clauses that arises in the arbitrary stage which will help us bound the survival probability. Nodes with same number of markings will appear consecutively in the controlled stage.

We will need the following definition:
\begin{definition}
A node $u$ with clause $C$ of width $w$ has \emph{effective width} $w'$ if it has $w - w'$ falsifying edges (as defined in \cref{subsubsec: constructing the tree}) arising out of it.
\end{definition}

Recall from \cref{subsubsec: constructing the tree} that a falsifying edge leads to a falsifying leaf, i.e. a node $v$ for which $F/v$ contains an empty clause. Therefore the effective width of a node corresponds to the number of children whose subtrees need to be explored.

Our controlled stage is divided into two substages:
\begin{enumerate}
\item[$\stage_1$.] $1$-marked nodes.
\item[$\stage_2$.] $2$-marked nodes corresponding to clauses of effective width $2$.
\end{enumerate}

\begin{remark}
  In stage $\stage_2$, an effective width $2$ (monotone) clause $C$ corresponds to a width $3$ (monotone) clause where one of the variables, say $x$, in $C$ will cause a falsifying edge.
  We will show that there exists a monotone clause $C'$ s.t. $x\in C'$ and remaining two literals in $C'$ are set to $1$.
  Since $F$ is negation-closed, the negation of $C'$ will have both its literals set to $0$, simplifying it to the clause $\neg x$. So, any node expanded using $C$ will have the edge labelled with $x$ as a falsifying edge.
\end{remark}

We first introduce useful notation regarding the disjoint stage. We then provide a detailed construction of each of the substages in the controlled stage followed by the impact of the controlled stage on the arbitrary stage.

\subsection{Additional notation for the disjoint stage}

As described in \cref{subsec: the disjoint stage}, we select a pseudomaximum size disjoint collection of clauses $C^0_0, C^0_1, \ldots, C^0_{t_0-1}$ in this stage.
Our analysis in \cref{sec: bound for small t0} will focus on obtaining an upper bound on $\total(u_0)$ where $u_0$ is an arbitrary node at level $t_0$. We will fix such a node $u_0$ for the rest of the section.
Write $C^0_i=\{p_i, x_i, x'_i\}$ where the variable $p_i$ is the label of the edge along the path $P(r\patharrow u_0)$. Let $V_0= \{0, 1, \ldots, t_0-1\}$.
For $i\in [t_0]$, let $X_i = \{x_i, x'_i\}$ and let $X = \cup_{i\in [t_0]} X_i$.
We will also need the following useful fact:

\begin{fact}\label{fact: after level t0 will always have one intersection}
Let $u$ be a node at level $\ell \ge t_0$ and let $C$ be a width $3$ monotone clause used to expand at $u$. Then, there must exist $i\in V_0$ such that $X_i\cap C\ne \emptyset$.
\end{fact}

\begin{proof}
Indeed, assume such $C$ existed. Then, $C$ must be disjoint from all clauses in $\cC^0$. However, this contradicts the fact $\cC^0$ is a pseudomaximum collection of clauses.
\end{proof}

\subsection{Stage \texorpdfstring{$\stage_1$}{1}: \texorpdfstring{$1$}{1}-marked nodes}

Let $F_1$ be the set of width $3$ monotone clauses that are live at $u_0$ and have exactly one marked variable at $u_0$. We select a pseudomaximum size disjoint collection $\cC^1$ of clauses from $F_1$ and expand $u_0$ to a sub-tree of uniform depth $t_1 = |\cC^1|$.

\begin{remark}\label{remark: maximum size cc1}
Similar to \cref{remark: maximum size cc0}, in the algorithmic implementation we will maintain an auxiliary data structure that will maintain the pseudomaximum collection $\cC^1$. Initially, the collection will be a maximal set of disjoint clauses from $F_1$ and whenever we come across a claim that uses maximum size property of $\cC^1$ and is violated, we will reset $\cC^1$ to the disjoint clauses supplied by the proof and `reset' this phase of the algorithm. Again, since a reset can happen at most $n$ times per $u_0$, this will increase the total runtime of the algorithm by factor of $n$. 
\end{remark}

Clauses in $F_1$ satisfy the following property:

\begin{fact}
\label{fact:disjoint}
If $C = \{ x_i, a,b\}$ and $C' =\{x'_i, c,d\}$ are in $F_1$ where $i\in V_0$,
then $\{a,b\} \cap \{c,d\} \neq \emptyset$.
\end{fact}

\begin{proof}
If $C$ and $C'$ are disjoint, we can replace the clause $C_i^0$ in $\cC^0$ with $C$ and $C'$ to obtain a larger size collection violating the pseudomaximum size condition on $\cC^0$.
\end{proof}

Let $V_1 = \{i \mid \text{a variable from $X_i$ appears in a clause from $\cC^1$}\}$.
Note that $|V_1| = |\cC^1| = t_1$.
By \cref{fact: after level t0 will always have one intersection}, for every $C\in F_1$, there is a unique $i\in V_0$ such that $C$ contains exactly one of $x_i$ or $x'_i$ with a single marking, and the remaining two variables in $C$ do not appear in any clause from $\cC^0$. 
Furthermore, by \cref{fact:disjoint}, for each $i\in V_0$, at most one variable from $X_i$ can appear in some clause of $\cC^1$; that is, if $x_i$ ($x'_i$) appears in some clause of $\cC^1$,  then $x'_i$ ($x_i$) does not appear in any other clause of $\cC^1$. These two indeed imply that $|V_1| = |\cC^1| = t_1$. 

We index the clauses in $\cC^1$ using $V_1$.
So, for $i\in V_1$, we write clause $C^1_i$ as $\{\tilde{x}_i, y_i, y'_i\}$ where $\tilde{x}_i\in X_i$.
For $i\in V_1$, let $Y_i = \{y_i, y'_i\}$.
Let $V_B = V_0 - V_1$ and $m_B = |V_B| = t_0 - t_1$.  

Before we describe stage $\stage_2$, we make some careful observations to prepare for it.

\subsection{Preparation for Stage \texorpdfstring{$\stage_2$}{2}}

For a node $u$ at the end of the stage $\stage_1$, define $F_{2}(u)$ to be the set of monotone width $3$ clauses $C$ that are live at $u$ and have exactly two singly marked variables (so that $C$ has mass $2$).

\begin{lemma}
\label{lemma:structure}
Let $u$ be an arbitrary node at the end of stage $\stage_1$.
Let $C\in F_2(u)$ be arbitrary.
Then, $C$ must be of one of the following forms:
\begin{enumerate}
    \item $C =\{\tilde{x}_i, \tilde{y}_i, z\}$  
    for some $i\in V_1, \tilde{x}_i\in X_i\setminus C^0_i, \tilde{y}_i\in Y_i$ and where $z$ does not appear in the shoot $S(r\patharrow u)$.
    \item $C =\{\tilde{x}_i, \tilde{x}_j, z\}$ for some  $i\in V_1, j\in V_B, \tilde{x}_i\in X_i\setminus C^0_i, \tilde{x}_j\in X_j$ and $z$ does not appear in the shoot $S(r\patharrow u)$.
    \item $\{\tilde{x}_i, \tilde{x}_j, z\}$ where $i, j \in V_B, \tilde{x}_i\in X_i, \tilde{x}_j\in X_j$ and $z$ does not appear in the shoot $S(r\patharrow u)$.
    \item 
    $\{\tilde{x}_i, \tilde{y}_j, z\}$ where $i\in V_B, j\in V_1, \tilde{x}_i\in X_i, \tilde{y}_j\in Y_j$, and $z$ does not appear in the shoot $S(r\patharrow u)$.
\end{enumerate}
\end{lemma}

\begin{proof}
Since $C$ has mass $2$, there must be $z\in C$ that does not appear in the shoot $S(r\patharrow u)$.
By \cref{fact: after level t0 will always have one intersection}, there exists $i\in V_0$ such that $X_i\cap C\ne \emptyset$.
We take cases on whether $i\in V_1$ or not and whether there exist two variables in $C$ from $X$.
\begin{enumerate}
    \item[Case 1.] $i\in V_1$. 
    Since $C\in F_2(u)$ and $\tilde{x}_i$ is singly marked, $\tilde{x}_i\not\in C^1_i\cap X_i$.
    
    Assume that the second marked variable in $C$ is from $X$.
    By \cref{fact:disjoint}, the second marked variable must be from $X_j$ where $j\in V_0$ and $j\ne i$.
    We claim that $j\in V_B$. Assume for contradiction that $j\in V_1$.
    Since $\tilde{x}_j$ is singly marked, $\tilde{x}_j\not\in C^1_j$.
    Then $C^1_i, C^1_j$ and $C$ are disjoint clauses that we can use to replace clauses $C^0_i, C^0_j$ from $\cC^0$ and obtain a larger set of disjoint clauses, contradicting the fact that $\cC^0$ is a pseudomaximum collection of clauses. Hence, the claim follows.
    
    If the second marked variable in $C$ is not from $X$, then it must be from $Y$.
    Let $\tilde{y}_j\in C$ where $j\in V_1, \tilde{y}_j\in Y_j$.
    We claim that $i=j$ in this case. Indeed, if not then the clauses $C^1_i$ and $C$ will be disjoint clauses containing distinct variables from $X_i$, violating \cref{fact:disjoint}. Hence, the claim follows.
    
    \item[Case 2.] $i\not\in V_1$. 
    As $V_B = V_0\setminus V_1$, we infer that $i\in V_B$.
    In this case, if the second marked variable in $C$ is from $X$, it must be from $X_j$ where $j\in V_B$ (if $j\in V_1$, then we are in the previous case) and the claim follows.
    Otherwise, the second marked variable in $C$ is from $Y$ and the claim follows as well.
\end{enumerate}
\end{proof}

Let $u^*$ denote the unique node at the end of stage $\stage_1$ where the path $P(u_0 \patharrow u^*)$ consists of only marked edges, each of which is labeled by a variable from $X_i$ for $i \in V_1$.
We note a useful relationship between the  live clauses at an arbitrary node $u$ at the end of $\stage_1$ and the live clauses at  $u^*$:

\begin{fact}
\label{fact:f2}
    For every node $u$ at the end of stage $\stage_1$, $F_{2}(u) \subseteq F_{2}(u^*)$.
\end{fact}

\begin{proof}
Let $C\in F_2(u)$ be arbitrary.
If $C$ is live at $u^*$, then $C$ must have exactly $2$ markings since $S(r, u) = S(r, u^*)$ and the markings only depend on the shoot.
Hence, we show that all such $C$ are live at $u^*$.
Since $C$ is live at $u$, it is also live at $u_0$.
Hence, if $C$ is not live at $u^*$, then it must be that $C$ contains one of the variables from $P(u_0, u^*)$. 
Equivalently, $C$ must have to contain $\tilde{x}_i$ where $i\in V_1$ and $\tilde{x}_i\in C^0_i$.
This cannot happen since by \cref{lemma:structure}, it must be that $\tilde{x}_i\in X_i\setminus C_i^0$.
\end{proof}

For simplicity we write $F_2 := F_{2}(u^*)$.
Let $F_{2R} = \{C\in F_2: \tilde{x}_i\in C \textrm{ where } i\in V_1\}$.
Let $F_{2B} = F_2 \setminus F_{2R}$.

\begin{fact}\label{fact:maximal disjoint collection size from F20}
Any maximally disjoint set of clauses from $F_{2B}$ has size at most $m_B$.
\end{fact}

\begin{proof}
By choice of $F_{2B}$, it only contains clauses of type $3$ or type $4$ as laid out in \cref{lemma:structure}.
Assume that there exists a collection $S$ of more than $m_B$ disjoint clauses from $F_{2B}$.
Let $T = \{C_i^0: i\in V_1\}$. We see that clauses in $S$ are disjoint from $T$, and $S$ and $T$ have no clauses in common.
However then, $S\cup T$ is a disjoint collection of clauses of size 
\[
|S| + |T| \ge (m_B + 1) + |V_1| = |V_0| + 1 = t_0 + 1
\]
violating the fact that $\cC^0$ is a pseudomaximum collection of clauses.
\end{proof}

\begin{remark}
We use this fact later in \cref{lemma:heavy}. So even though this fact does not algorithmically provide us with clauses to replace $\cC^0$ with, for algorithm's sake, we only care about the maximal collection we encounter from \cref{lemma:heavy} and there indeed, we can constructively find such a collection if pseudomaximum property is violated.
\end{remark}

\begin{fact}\label{fact: at end of stage 1 live clauses from F2R}
Let $C\in F_{2R}$ be arbitrary. Let $i\in V_1$ be such that $\tilde{x}_i\in C$. Let $u$ be arbitrary node at the end of stage $\stage_1$ where a variable from $X_i$ appears along the path $P(u_0\patharrow u)$. 
Then $C$ is live at $u$.
\end{fact}

\begin{proof}
By \cref{lemma:structure}, $C$ can only take one of two forms:
If $C$ is of the form $(\tilde{x}_i, \tilde{x}_j, z)$ where $j\in V_B$ and $z$ does not appear along the shoot $S(u_0\patharrow u)$, then this follows.
Otherwise, $C$ must be of the form $(\tilde{x}_i, \tilde{y}_i, z)$ where $z$ is not along the shoot $S(u_0\patharrow u)$ and $\tilde{y}_i\in Y_i$.
By assumption, a variable from $X_i$ is in the path $P(u_0\patharrow u)$.
This can only happen at the node corresponding to the clause $C^1_i$.
As $\tilde{y}_i$ appears exactly once along the shoot $S(u_0\patharrow u)$ - at the node corresponding to the clause $C^1_i$ - we infer that $\tilde{y}_i$ is not along the path $P(u_0\patharrow u)$.
Hence, the clause $C$ is still live at $u$.
\end{proof}

Let $V_R =\{i \in V_1 \mid \exists C\in F_{2R}
\text{ such that } X_i \cap C \ne \emptyset\}$.
Let $m_R =|V_R|$.
Let $m_I = |V_1 \setminus V_R|$.
We have $m_B+m_R+m_I = t_0 = \frac{n}{4} - \Delta$. Fix a pseudomaximum size collection  $\cC'_R$ of disjoint clauses from $F_{2R}$. 
Let $V'_R = \{ i\in V_R \mid x'_i \text{ appears in a clause in $\cC_R'$}\}$ and $m'_R = |V'_R| = |\cC'_R|$.
Observe that $m'_R \le m_R$.

\begin{remark}
Similar to \cref{remark: maximum size cc0} and \cref{remark: maximum size cc1}, in the algorithmic implementation, we will maintain an auxiliary data structure that will maintain $\cC'_R$ to be a pseudomaximum set of disjoint clauses from $F_{2R}$ and whenever we come across a claim that uses this property but is violated, we will reset $\cC'_R$ to be the disjoint clauses supplied by the proof and will reset stage $\stage_1$. Such reset can happen at most $n$ times per $u_0$ and hence, the runtime of the algorithm can be increased by at most $n$. We will allow remaining claims in this section to use that $\cC'_R$ is a maximum sized collection.
\end{remark}

\subsection{Stage \texorpdfstring{$\stage_2$}{2}: \texorpdfstring{$2$}{2}-marked nodes with effective width 2 clauses}

Consider a node $u$ at the end of the stage $\stage_1$.  Let $V^{\textrm{marked}}(u)$ denote the set of marked variables along the root to $u$ path $P(r\patharrow u)$.
Let $\cC'_{R}(u) = 
\{C\in \cC'_R\mid \exists i \in V^{\textrm{marked}}(u) \text{ such that } C\cap X_i\ne \emptyset\}$.
Let $\ell(u) = |\cC'_{R}(u)| =|V^{\textrm{marked}}(u) \cap V'_R$|.

In stage $\stage_2$, we expand using the clauses in $\cC'_{R}(u)$.
By \cref{fact: at end of stage 1 live clauses from F2R}, these clauses are live at $u$ and disjoint from each other,
the expansion will be carried out for exactly $\ell(u)$ levels
where each clause in $\cC'_{R}(u)$ will be used to expand one level.
We record this fact here:
\begin{fact}\label{fact: length of stage A2}
Let $u$ be a node at the end of stage $\stage_1$. Then, stage $\stage_2$ underneath $u$ has length $\ell(u) = |\cC'_{R}(u)| =|V^{\textrm{marked}}(u) \cap V'_R|$
\end{fact}

After this expansion, we finish the controlled stage and enter the arbitrary stage. We record the following useful property of nodes in this stage:

\begin{lemma}\label{lemma: stage 1(c) has length l(u)}
All nodes in stage $\stage_2$ will have effective width $2$ and mass at most $\frac{3}{2}$. 
\end{lemma}

\begin{proof}
Let $v$ be arbitrary node in stage $\stage_2$ developed using clause $C$ where $C\in \cC'_R(u)$.
Let $i\in V^{\textrm{marked}}(u)$ be such that $\tilde{x}_i\in C$.
As $C^0_i = (p_i \lor x_i \lor x'_i)$ is a clause in $\cC^0$, and $F$ is negation-closed, the negation-clause of $C^0_i$ is also in $F$.
At $v$, the negation-clause of $C^0_i$ simplifies to the unit clause $\neg \tilde{x}_i$.
So, at $v$, the edge with label $\tilde{x}_i$ will be a falsifying edge, making its effective width equal to $2$.
Moreover, since $v$ is $2$-marked, some other edge from $C$ is also marked.
This implies the mass of $v$ will indeed be at most $\frac{3}{2}$ as desired.
\end{proof}

\subsection{Arbitrary Stage}

Let $u$ be arbitrary node at the end of the controlled stage.
In the controlled stage, we expand using any monotone clause that is available and put no restrictions. However, we will still be able to argue regarding the kinds of clauses that one could encounter in this stage.

We begin by showing that every $1$-marked vertex in this stage will have mass at most $\frac{9}{4}$:

\begin{lemma}
\label{lemma:phase 1 double marking has small mass}
Every $1$-marked node in arbitrary stage has mass at most $\frac{9}{4}$.
\end{lemma}

\begin{proof}
Indeed, if a $1$-marked vertex $v$ with clause $C$ in arbitrary stage has a larger mass, then it must have mass $\frac{5}{2}$.
In that case, there exists a unique $i\in V_0$ such that $C$ contains exactly one element from $X_i$. Since mass of $v$ is $\frac{5}{2}$ and $i\not\in V_1$, the remaining two variables do not appear in the shoot $S(r\patharrow u)$.
However, this implies $C$ is disjoint from $\cC^1$, contradicting the fact that $\cC^1$ is a pseudomaximum disjoint set of clauses with single marking.
\end{proof}

\begin{lemma}
\label{lemma:heavy}
For any root to leaf shoot $S$ in $T(u)$, the number of nodes with $2$ marked edges and mass $2$ is at most $m'_R+m_B-\ell(u)$.
\end{lemma}

\begin{proof}
Call such clauses as \emph{heavy clauses}. By \cref{fact:f2}, heavy clauses along any shoot during arbitrary stage are all in $F_2$. Observe that these heavy clauses must be disjoint from $\cC'_{R}(u)$. We first claim that there can be at most $m'_R-\ell(u)$ heavy clauses from $F_{2R}$ during arbitrary stage along $S$. Indeed, if there were more, then these heavy clauses combined with $\cC'_R(u)$ would form a collection of disjoint clauses of size at least $(m'_R - \ell(u) + 1) + \ell(u) = m'_R + 1$, violating the fact that $\cC'_R$ is a pseudomaximum collection of disjoint clauses.

The only other heavy clauses that can occur in arbitrary stage are from the set $F_{2B}$. By \cref{fact:maximal disjoint collection size from F20}, there are at most $m_B$ disjoint clauses in $F_{2B}$. Since heavy clauses along a shoot must be disjoint from each other, there can be at most $m'_R + m_B - \ell(u)$ heavy clauses in $S$.
\end{proof}

\section{Bounding \texorpdfstring{$\total$}{psi} when \texorpdfstring{$t_0\leq \frac{n}{4}$}{t0 <= n/4}}\label{sec: bound for small t0}\label{sec: small t0}

In this section, we show that if in a transversal tree $T$, the number of maximally disjoint clauses chosen in the disjoint stage is at most $n/4$, then $\total(r) \le 6^{n/4}$ where $r$ is the root of $T$.
Formally, our main theorem is:

\begin{theorem}\label{thm: bound for small t0}
Let $T$ be a transversal tree developed using the clause selection criteria as laid out in \cref{section:canonical} and $\cref{sec: controlled}$ with $t_0 \le \frac{n}{4}$.
Then, $\total(r)\le 6^{\frac{n}{4}}$ where $r$ is the root of $T$.
\end{theorem}

To bound $\total(r)$, we will carefully count and sum up the survival values of all nodes that appear at the end of the disjoint stage.
To facilitate that, we need a handle on the survival value of a subtree that is developed in arbitrary stage. We introduce the following quantity to help us with that:

Let $M(w, d, h) = \max_{u} \total(u)$ where $u$ is a node at depth $\frac{n}{2}-d$ in arbitrary stage, every root to leaf shoot in $T(u)$ has weight at least $w$, and for every root to leaf shoot, the number of $2$-marked nodes with mass $2$ is bounded by $h$.

As $u$ is a node at level $\frac{n}{2}-d$, the depth of $T(u)$ is $d$. Moreover, as $u$ is in arbitrary stage, every node in $T$ has at least one marked edge coming out of it, and by \cref{lemma:phase 1 double marking has small mass}, every $1$-marked node has that marked edge with survival probability at most $1/4$.


We will define the following useful function:

\begin{align*}
F(w,d,h) &=
\begin{cases}
\left(\frac{9}{4}\right)^d & \text{ if $w\le d$}\\
\left(\frac{9}{4}\right)^{2d-w}2^{w-d} & \text{ if $d\le w\le d+h$}\\
\left(\frac{9}{4}\right)^{2d-w}2^{h}\left(\frac{27}{8}\right)^{(w-d-h)/2} 
= 2^{h}\left(\frac{27}{8}\right)^{(3d-w-h)/2}\left(\frac{3}{2}\right)^{w-2d}
& \text{ if $d+h\le w\le 3d-h$}\\
2^{3d-w}\left(\frac{3}{2}\right)^{w-2d} & \text{ if $3d-h\le w$}\\
\end{cases}
\end{align*}

We will use the following bound on $M(w, d, h)$ that we prove in \cref{subsec: appendix proof bound in stage b}.

\begin{lemma}\label{lemma: bounding Mwdh by Fwdh}
For all $w, d, h$: $M(w, d, h) \le F(w, d, h)$.
\end{lemma}

With this, we are ready to prove \cref{thm: bound for small t0}:

\begin{proof}[Proof of \cref{thm: bound for small t0}]

Let $u_0\in T$ be an arbitrary node at depth $t_0$.
Then, we can associate quantities $m_B(u_0), m'_R(u_0), t_1(u_0)$ with the subtree $T(u_0)$.
We bound $\total(u_0)$ in terms of these quantities and function $F$ from \cref{lemma: bounding Mwdh by Fwdh}.
We will sum over $\total(u)$ where $u$ is a node at the end of controlled stage and then use $F$ to bound $\total(u)$ as $u$ will be at the beginning of arbitrary stage. We will also need to precisely compute $\sigma(u_0, u)$, for such $u$ and for that, we keep track of how many marked edges (say $i$) from stage $\stage_1$ corresponding to $m'_R(u_0)$ are on the path from $u_0$ to $u$.
This $i$ will also be the length of stage $\stage_2$, which will further help us in bounding $\sigma(u_0, u)$.

To do that, we first introduce the following quantities that we will use in the upper bound:
Let $w(u_0, i) = \frac{n}{2} - 2i - t_1(u_0), d(u_0, i) = \frac{n}{2} - t_0 - t_1(u_0) - i, h(u_0, i) = m'_R(u_0) + m_B(u_0) - i$.
Using these quantities, we will upper bound $\total(u_0)$ using the following expression:
Define
\[
N(u_0) = \left(\frac{5}{2}\right)^{t_1(u_0)} \left(\frac{4}{5}\right)^{m'_R(u_0)}\sum_{i=0}^{m'_R(u_0)} \binom{m'_R(u_0)}{i}\left(\frac{3}{8}\right)^{i}\cdot F(w(u_0, i), d(u_0, i), h(u_0, i))
\]

Using this function, we will obtain the following as our main lemma:

\begin{lemma}\label{lem: sigma T u0 is less than N u0}
$\total(u_0) \le N(u_0)$.
\end{lemma}

We assume this bound holds and continue our proof of \cref{thm: bound for small t0}. We will prove the lemma in \cref{subsec: proving the main lemma}.

As $N(u_0)$ depends on $F$, we want to figure out what case for the function of $F$ applies. Recall that this depends on the relationship between $w(u_0, i), d(u_0, i), h(u_0, i)$. To help with this, we introduce the following function:
\[
    I(u_0) = 3t_0 + 2t_1(u_0) + m'_R(u_0) + m_B(u_0)
\]

We show that the values of $I$ and $n$ govern which of the $4$ functions will $F$ equal. This is surprising since $I$ is a function of $u_0$ and not a function of $i$. We first show that only $2$ function choices of $F$ can arise. We do this by showing:

\begin{claim}\label{claim: d + h le w}
$d(u_0, i) + h(u_0, i)\le w(u_0, i)$.
\end{claim}

\begin{proof}[Proof of \cref{claim: d + h le w}]
Indeed we compute:     
\begin{align*}
d(u_0, i) + h(u_0, i)
& = \frac{n}{2} - t_0 - t_1(u_0) + m'_R(u_0) + m_B(u_0) - 2i\\
& = w(u_0, i) + (m'_R(u_0) + m_B(u_0) - t_0)\\
& \le w(u_0, i)
\end{align*}
where the last inequality follows because $m'_R(u_0) + m_B(u_0)\le t_1(u_0) + m_B(u_0) = t_0$.
\end{proof}

We next show that the value of $I$ decides the choice function for $F$:

\begin{claim}\label{claim: w le 3d-h iff I le n}
$w(u_0, i)\le 3d(u_0, i) - h(u_0, i)$ if and only if $I(u_0)\le n$.
\end{claim}

\begin{proof}[Proof of \cref{claim: w le 3d-h iff I le n}]
We compute the following:
\begin{align*}
3d(u_0, i) - h(u_0, i)
& = \frac{3n}{2} - 3t_0 - 3t_1(u_0) - m'_R(u_0) - m_B(u_0) - 2i\\
& = w(u_0, i) + n - 3t_0 - 2t_1(u_0) - m'_R(u_0) - m_B(u_0)
\end{align*}
and hence, 
\[
w(u_0, i)\le 3d(u_0, i) - h(u_0, i)
\iff
3t_0 + 2t_1(u_0) + m'_R(u_0) + m_B(u_0) = I(u_0) \le n
\]
\end{proof}

With this, we bound $\total(r)$ as follows:
\begin{align*}
\total(r)
& = \sum_{u_0: u_0 \textrm{ at depth } t_0} \total(u_0)\\
& \le \sum_{u_0: u_0 \textrm{ at depth } t_0} N(u_0)\\
& = \sum_{u_0: u_0 \textrm{ at depth } t_0, I(u_0)\le n} N(u_0) + \sum_{u_0: u_0 \textrm{ at depth } t_0, I(u_0) > n} N(u_0)\\
& \le 3^{t_0}\max\left( \max_{u_0: I(u_0)\le n} N(u_0), \max_{u_0: I(u_0) > n} N(u_0) \right)\\
& = \max\left(\max_{u_0: I(u_0)\le n} 3^{t_0}\cdot N(u_0), \max_{u_0: I(u_0)\ge n} 3^{t_0}\cdot N(u_0) \right)\\
\end{align*}

We show that the inner quantities are maximized when $I(u_0) = n$ by the following two claims. These claims just rely on the inequalities listed in the claim, proving that some expression is bounded. We see these claims as solving an optimization problem and do not rely on any properties of the transversal tree.

\begin{claim}\label{claim: optimization: when I le n max when I is n}
Let $u_0$ be such that $m_B(u_0) + t_1(u_0)\le t_0, m'_R(u_0)\le t_1(u_0), I(u_0)\le n$.
Then, $3^{t_0}\cdot N(u_0)$ is maximised when $I(u_0) = n$.
\end{claim}

\begin{claim}\label{claim: optimization: when I ge n max when I is n}
Let $u_0$ be such that $m_B(u_0) + t_1(u_0)\le t_0, m'_R(u_0)\le t_1(u_0), I(u_0)\ge n$.
Then, $3^{t_0}\cdot N(u_0)$ is maximised when $I(u_0) = n$.
\end{claim}

Lastly, we show that when $I(u_0) = n$, then the inner quantity is bounded by $6^{n/4}$:

\begin{claim}\label{claim: optimizaion: when I is n bound 6 to n over 4}
Let $u_0$ be such that $m_B(u_0) + t_1(u_0)\le t_0, m'_R(u_0)\le t_1(u_0), I(u_0) = n$.
Then, $3^{t_0}\cdot N(u_0)\le 6^{n/4}$.
\end{claim}

These $3$ claims together indeed show that $\total(r) \le 6^{n/4}$ as desired. We defer the proofs of all these claims to \cref{subsec: appendix proof claims from thm: bound for small t0}.

\end{proof}

\section{Conclusion}
We gave an optimal algorithm for the Not-All-Equal variant of $\enumballsat{k}{\frac{n}{2}}$ for $k = 3$. Extending the analysis of our algorithm to large $k$ would break SSETH. However, extending this to even $k = 4$ poses a great challenge.

\section{Acknowledgments}

We want to thank Pavel Pudl{\'a}k for helpful discussions.


\printbibliography

\appendix

\section{Deferred proofs}

\subsection{Remaining proofs from \texorpdfstring{$t_0 \ge n/4$}{t0 >= n/4}}\label{subsec: appendix proof bounding mwd with fwd}

We prove the main lemma from \cref{sec: bound for large t0}:

\begin{proof}[Proof of \cref{lem: bounding mwd with fwd}]
Let $M(w, d) = \max_{u} \total(u)$. Note that as $u$ appears at level $\frac{n}{2}-d$, the depth of $T(u)$ is $d$. Moreover, as $u$ appears at level $\ge t_0$, by \cref{fact: after level t0 will always have marking}, every node in $T(u)$ has at least one marked edge coming out of it.
Let edges out of $u$ be $e_1, e_2, e_3$ and the nodes $v_1, v_2, v_3$.
Note that the event that say marked edge $e_1$ is superfluous is independent of the fact that $u$ survives.
By considering cases on the number of markings at the root of a transversal tree, we see that $M(w, d)$ obeys the following recurrence (we don't include the case where there is a falsifying edge or that number of children are less than $3$ since that case is dominated by the cases below):
\begin{align*}
M(w, d) =     
\max\{
& (5/2) M(w-1, d-1)\\
& 2 M(w-2, d-1)\\
& (3/2) M(w-3, d-1)\\
\}\\
\end{align*}

with the base cases defined as follows:

\[
M(w, 0) =
\begin{cases}
1 & w \le 0\\
0 & w > 0
\end{cases}
\]

Let $G_1(w, d) = (\frac{5}{2})^{2d-w} 2^{w-d}$ and let $G_2(w, d) = 2^{3d-w} (\frac{3}{2})^{w-2d}$.
Let $G(w, d) = \min(G_1(w, d), G_2(w, d))$.

The result follows from the following claims:

\begin{claim}\label{claim: t0 > n/4: G = F}
$G(w, d) = F(w, d)$.
\end{claim}

\begin{claim}\label{claim: t0 > n/4: M <= G1}
$M(w, d) \le G_1(w, d)$.
\end{claim}

\begin{claim}\label{claim: t0 > n/4: M <= G2}
$M(w, d) \le G_2(w, d)$.
\end{claim}

Indeed once we prove these claims, we can conclude that when $w\le 2d$, then $M(w, d)\le G_1(w, d) = G(w, d) = F(w, d)$ and similarly, when $w\ge 2d$, we conclude $M(w, d)\le G_2(w, d) = G(w, d) = F(w, d)$.
Hence, for all $w, d$, we conclude that $M(w, d) \le F(w, d)$ as desired.

We now prove these claims one by one:
\begin{proof}[Proof of \cref{claim: t0 > n/4: G = F}]
We compute that 
\[
\frac{G_1(w, d)}{G_2(w, d)}
= \left(\frac{5}{2}\right)^{2d-w} 2^{2w-4d} \left(\frac{3}{2}\right)^{2d-w}
= \left(\frac{15}{16}\right)^{2d-w}
\]
Hence, $\frac{G_1(w, d)}{G_2(w, d)}\le 1$ if and only if $w\le 2d$.
This implies that when $w\le 2d$, $G(w, d) = G_1(w, d)$ and when $2d\le w$, $G(w, d) = G_2(w, d)$.
We compare this to the definition of $F(w, d)$ and conclude that indeed $F(w, d) = G(w, d)$ as desired.
\end{proof}

\begin{proof}[Proof of \cref{claim: t0 > n/4: M <= G1}]
We proceed by induction on $d$.
When $d = 0$, the claim trivially holds.
For $d\ge 1$, we compute:
\begin{align*}
M(w, d)
& = \max((5/2)M(w-1, d-1), 2M(w-2, d-1), (3/2)M(w-3, d-1))\\
& \le \max((5/2)G_1(w-1, d-1), 2G_1(w-2, d-1), (3/2)G_1(w-3, d-1))\\
& \le \max\left((5/2) \left(\frac{5}{2}\right)^{2d-w-1}2^{w-d}, 2\left(\frac{5}{2}\right)^{2d-w}2^{w-d-1}, (3/2)\left(\frac{5}{2}\right)^{2d+1}2^{w-d-2}\right)\\
& = \left(\frac{5}{2}\right)^{2d-w}2^{w-d} \max\left(1, 1, \frac{15}{16}\right)\\
& = \left(\frac{5}{2}\right)^{2d-w}2^{w-d}\\
& = G_1(w, d)
\end{align*}
\end{proof}

\begin{proof}[Proof of \cref{claim: t0 > n/4: M <= G2}]
We proceed by induction on $d$.
When $d = 0$, the claim trivially holds.
For $d\ge 1$, we compute:
\begin{align*}
M(w, d)
& = \max((5/2)M(w-1, d-1), 2M(w-2, d-1), (3/2)M(w-3, d-1))\\
& \le \max((5/2)G_2(w-1, d-1), 2G_2(w-2, d-1), (3/2)G_2(w-3, d-1))\\
& \le \max\left((5/2) \left(\frac{3}{2}\right)^{w-2d+1}2^{3d-w-2}, 2\left(\frac{3}{2}\right)^{2d-w}2^{3d-w-1}, (3/2)\left(\frac{3}{2}\right)^{w-2d}2^{3d-w-1}\right)\\
& = \left(\frac{3}{2}\right)^{w-2d}2^{3d-w} \max\left(\frac{15}{16}, 1, 1\right)\\
& = \left(\frac{3}{2}\right)^{w-2d}2^{3d-w}\\
& = G_2(w, d)
\end{align*}
\end{proof}

\end{proof}

\subsection{Remaining proofs from \texorpdfstring{$t_0 \le n/4$}{t0 <= n/4}}

We here prove the remaining proofs from \cref{sec: bound for small t0}.

\subsubsection{Proof of \texorpdfstring{\cref{lem: sigma T u0 is less than N u0}}{main lemma}}\label{subsec: proving the main lemma}

We lastly present proof showing conditional survival probability of subtree underneath arbitrary node $u_0$ at end of the disjoint stage is bounded by $N(u_0)$:

\begin{proof}[Proof of \cref{lem: sigma T u0 is less than N u0}]
\begin{align*}
& \total(u_0)\\
& = \sum_{u: u \textrm{ at end of controlled stage}} \sigma(u_0\patharrow u)\cdot \total(u)\\
& = \sum_{i=0}^{m'_R(u_0)}\sum_{\substack{u: u \textrm{ at end of controlled stage,}\\ u \textrm{ at depth } t_0 + t_1(u_0) + i}} \sigma(u_0\patharrow u)\cdot \total(u)\\
& \le \sum_{i=0}^{m'_R(u_0)}\sum_{\substack{u: u \textrm{ at end of controlled stage,}\\ u \textrm{ at depth } t_0 + t_1(u_0) + i}} \sigma(u_0\patharrow u)\cdot\\
& M\left(w = \frac{n}{2} - 2i - t_1(u_0), d = \frac{n}{2} - t_0 - t_1(u_0) - i, h = m'_R(u_0) + m_B(u_0) - i\right) \\
\end{align*}

Recall that $w(u_0, i) = \frac{n}{2} - 2i - t_1(u_0), d(u_0, i) = \frac{n}{2} - t_0 - t_1(u_0) - i, h = m'_R(u_0) + m_B(u_0) - i$.
We sum over nodes with exactly $i$ marked variables out of $m'_R(u_0)$ in stage $\stage_1$, i.e., over nodes $u$ where $\ell(u) = i$. Using \cref{fact: length of stage A2}, we infer that for such nodes, stage $\stage_2$ has length $i$ and hence, controlled stage ends at level $t_0 + t_1 (u_0) + i$.
The parameters $w(u_0, i), d(u_0, i)$ are indeed what are present above since every root to leaf shoot $S$ in the tree $T$ has $w(S) = d(S) = \frac{n}{2}$ by \cref{fact:shoot-extended-weight-lower-bound}. Our bound on $h(u_0, i)$ follows from \cref{lemma:heavy}.

We continue our computation:

\begin{align*}
& \total(u_0)\\
& \le \sum_{i=0}^{m'_R(u_0)}\sum_{\substack{u: u \textrm{ at end of controlled stage,}\\ u \textrm{ at depth } t_0 + t_1(u_0) + i}} \sigma(u_0\patharrow u)\cdot M\left(w(u_0, i), d(u_0, i), h(u_0, i)\right) \\
& \le \sum_{i=0}^{m'_R(u_0)}\sum_{\substack{u: u \textrm{ at end of controlled stage,}\\ u \textrm{ at depth } t_0 + t_1(u_0) + i}} \sigma(u_0\patharrow u)\cdot F\left(w(u_0, i), d(u_0, i), h(u_0, i)\right) \\
& \le \sum_{i=0}^{m'_R(u_0)}\sum_{\substack{u: u \textrm{ at end of controlled stage,}\\ u \textrm{ at depth } t_0 + t_1(u_0)}} \sigma(u_0\patharrow u) \left(\frac{3}{2}\right)^i\cdot F\left(w(u_0, i), d(u_0, i), h(u_0, i)\right)\\ 
\end{align*}

For the last inequality, we sum over all nodes at the last $i$ levels of controlled stage. These nodes are in stage $\stage_2$ and by \cref{lemma: stage 1(c) has length l(u)}, each of them has mass $\frac{3}{2}$.

We further simplify the sum by using the fact that each $1$-marked node in controlled stage has mass $\frac{5}{2}$, that number of paths with $i$ out of $m'_R(u_0)$ is $\binom{m'_R(u_0)}{i}\cdot 2^{m'_R(u) - i}$, those paths have survival probability $\left(\frac{1}{2}\right)^i$ due to these $i$ marked edges. Remaining $t - m'_R(u_0)$ nodes each have mass $\frac{5}{2}$ and don't affect the values $w(u_0 ,i), d(u_0 ,i), h(u_0 ,i)$. Hence, we end up with the following expression:

\begin{align*}
& \total(u_0)\\
& = \sum_{i=0}^{m'_R(u_0)} \binom{m'_R(u_0)}{i}\left(\frac{5}{2}\right)^{t_1(u_0) - m'_R(u_0)}\left(2\right)^{m'_R(u_0)-i}\left(\frac{1}{2}\right)^i\left(\frac{3}{2}\right)^i\cdot F\left(w(u_0, i), d(u_0, i), h(u_0, i)\right)\\
& = \left(\frac{5}{2}\right)^{t_1(u_0)} \left(\frac{4}{5}\right)^{m'_R(u_0)}\sum_{i=0}^{m'_R(u_0)} \binom{m'_R(u_0)}{i}\left(\frac{3}{8}\right)^{i}\cdot F\left(w(u_0, i), d(u_0, i), h(u_0, i)\right)\\
& = N(u_0)
\end{align*}
\end{proof}

\subsubsection{Various claims from \texorpdfstring{\cref{thm: bound for small t0}}{Theorem: Bounds for small t0}}\label{subsec: appendix proof claims from thm: bound for small t0}

We present all the deferred proofs from \cref{thm: bound for small t0} one after the other:

\begin{proof}[Proof of \cref{claim: optimization: when I le n max when I is n}]
\begin{align*}
& \left(3\right)^{t_0}\cdot N(u_0)\\
= & \left(3\right)^{t_0}\left(\frac{5}{2}\right)^{t_1(u_0)} \left(\frac{4}{5}\right)^{m'_R(u_0)}\sum_{i=0}^{m'_R(u_0)} \binom{m'_R(u_0)}{i}\left(\frac{3}{8}\right)^{i}\cdot F\left(w(u_0, i), d(u_0, i), h(u_0, i)\right)\\
= & \left(3\right)^{t_0}\left(\frac{5}{2}\right)^{t_1(u_0)} \left(\frac{4}{5}\right)^{m'_R(u_0)}\sum_{i=0}^{m'_R(u_0)} \binom{m'_R(u_0)}{i}\left(\frac{3}{8}\right)^{i}\cdot \left(\frac{9}{4}\right)^{2d(u_0, i) - w(u_0, i)} \left(2\right)^{h(u_0, i)}\left(\frac{27}{8}\right)^{(w(u_0, i) - d(u_0, i) - h(u_0, i)) / 2}\\
= & \left(3\right)^{t_0}\left(\frac{5}{2}\right)^{t_1(u_0)} \left(\frac{4}{5}\right)^{m'_R(u_0)}\sum_{i=0}^{m'_R(u_0)} \binom{m'_R(u_0)}{i}\left(\frac{3}{8}\right)^{i}\cdot\\
& \left(\frac{9}{4}\right)^{n/2 - 2t_0 - t_1(u_0)} \left(2\right)^{m._R(u_0) + m_B(u_0) - i}\left(\frac{3\sqrt{6}}{4}\right)^{t_0 - m'_R(u_0) - m_B(u_0)}\\
= & \left(\frac{4\sqrt{6}}{9}\right)^{t_0}\left(\frac{10}{9}\right)^{t_1(u_0)}\left(\frac{4\sqrt{6}}{9}\right)^{m_B(u_0)}\left(\frac{3}{2}\right)^n \left(\frac{16\sqrt{6}}{45}\right)^{m'_R(u_0)}\sum_{i=1}^{m'_R(u_0)}\binom{m'_R(u_0)}{i}\left(\frac{3}{16}\right)^i\\
= & \left(\frac{4\sqrt{6}}{9}\right)^{t_0}\left(\frac{10}{9}\right)^{t_1(u_0)}\left(\frac{4\sqrt{6}}{9}\right)^{m_B(u_0)}\left(\frac{3}{2}\right)^n \left(\frac{19\sqrt{6}}{45}\right)^{m'_R(u_0)}\\
\end{align*}
We see that $\frac{4\sqrt{6}}{9} \ge 1$.
If $I(u_0) = 3t_0 + 2t_1(u_0) + m'_R(u_0) + m_B(u_0) < n$, then we can increase $t_0$ so that $I(u_0) = n$ and doing that will only increase the value of $3^{t_0}\cdot N(u_0)$.
Hence, it suffices to consider the case of when $I(u_0) = n$.
\end{proof}

\begin{proof}[Proof of \cref{claim: optimization: when I ge n max when I is n}]
\begin{align*}
& \left(3\right)^{t_0}\cdot N(u_0)\\
= & \left(3\right)^{t_0}\left(\frac{5}{2}\right)^{t_1(u_0)} \left(\frac{4}{5}\right)^{m'_R(u_0)}\sum_{i=0}^{m'_R(u_0)} \binom{m'_R(u_0)}{i}\left(\frac{3}{8}\right)^{i}\cdot F\left(w(u_0, i), d(u_0, i), h(u_0, i)\right)\\
= & \left(3\right)^{t_0}\left(\frac{5}{2}\right)^{t_1(u_0)} \left(\frac{4}{5}\right)^{m'_R(u_0)}\sum_{i=0}^{m'_R(u_0)} \binom{m'_R(u_0)}{i}\left(\frac{3}{8}\right)^{i}\cdot \left(2\right)^{3d(u_0, i) - w(u_0, i)}  \left(\frac{3}{2}\right)^{w(u_0, i) - 2d(u_0, i)}\\
= & \left(\frac{27}{32}\right)^{t_0}\left(\frac{15}{16}\right)^{t_1(u_0)} \left(\frac{4}{5}\right)^{m'_R(u_0)} \left(\frac{8}{3}\right)^{n/2}\sum_{i=0}^{m'_R(u_0)} \binom{m'_R(u_0)}{i}\left(\frac{3}{16}\right)^{i}\\
= & \left(\frac{27}{32}\right)^{t_0}\left(\frac{15}{16}\right)^{t_1(u_0)} \left(\frac{19}{20}\right)^{m'_R(u_0)} \left(\frac{8}{3}\right)^{n/2}\\
\end{align*}
We see that $\frac{27}{32} \le 1$.
If $I(u_0) = 3t_0 + 2t_1(u_0) + m'_R(u_0) + m_B(u_0) > n$, then we can decrease $t_0$ so that $I(u_0) = n$ and doing that will only increase the value of $3^{t_0}\cdot N(u_0)$.
Hence, it suffices to consider the case of when $I(u_0) = n$.
\end{proof}

\begin{proof}[Proof of \cref{claim: optimizaion: when I is n bound 6 to n over 4}]
Assume that $t_0, t_1(u_0), m'_R(u_0), m_B(u_0)$ are chosen to maximise the quantity  $\left(3\right)^{t_0}\cdot N(u_0)$ and they attain the maximum value possible.
We easily compute that
\begin{align*}
& \left(3\right)^{t_0}\cdot N(u_0)\\
= & \left(\frac{27}{32}\right)^{t_0}\left(\frac{15}{16}\right)^{t_1(u_0)} \left(\frac{19}{20}\right)^{m'_R(u_0)} \left(\frac{8}{3}\right)^{n/2}\\
= & \left(\frac{3\sqrt{6}}{2}\right)^{t_0}\left(\frac{5}{2}\right)^{t_1(u_0)} \left(\frac{19\sqrt{6}}{30}\right)^{m'_R(u_0)} \left(\frac{2\sqrt{6}}{3}\right)^{m_B(u_0)}\\
\end{align*}
We now take two cases:
\begin{enumerate}
\item[Case 1.]
$m'_R(u_0) / 2 \ge t_0 - (m_B(u_0) - t_1(u_0))$.
Let $x = t_0 - (m_B(u_0) - t_1(u_0))$.
Now, increase $t_1$ by $x$ and decrease $m'_R$ by $2x$.
As $\frac{5}{2} \ge \left(\frac{19\sqrt{6}}{30}\right)^2$, such a transformation would only increase the value of $3^{t_0}\cdot N(u_0)$.
This transformation still maintains the invariant that $I(u_0) = n$ and also, $m'_R(u_0)\le t_1(u_0)$.
In the resultant transformation, $m_B(u_0) + t_1(u_0) = t_0$.
Since our chosen values are optimal, we assume this is already the case.
We substitute for $m_B(u_0)$ and infer that
\[
3^{t_0}N(u_0)\le 6^{t_0} \left(\frac{5\sqrt{6}}{8}\right)^{t_1(u_0)}\left(\frac{19\sqrt{6}}{30}\right)^{m'_R(u_0)}
\]
with the inequalities in the remaining variables being: $4t_0 + t_1(u_0) + m'_R(u_0) = n$ and $m'_R(u_0)\le t_1(u_0)$.

As $\frac{5\sqrt{6}}{8} < \frac{19\sqrt{6}}{30}$, if $t_1(u_0) - m'_R(u_0) = x$, then to get a larger value, we can increase $m'_R(u_0)$ by $x/2$ and decrease $t_1(u_0)$ by $x/2$.
This still maintains the invariant $4t_0 + t_1(u_0) + m'_R(u_0) = n$ and then, $m'_R(u_0) = t_1(u_0)$.
Since our chosen values are optimal, we assume this is already the case.
We substitute for $m'_R(u_0)$ and infer that:
\[
3^{t_0}N(u_0)\le 6^{t_0} \left(\frac{19}{8}\right)^{t_1(u_0)}
\]
where the only constrain is $4t_0 + 2t_1(u_0) = n$.

As $6 > \left(\frac{19}{8}\right)^2$, if $t_1(u_0) = x$, then the transformation where we decrease $t_1(u_0)$ by $x$ and increase $t_0$ by $x/2$ maintains the invariant $4t_0 + 2t_1(u_0) = n$ and the value of $3^{t_0}N(u_0)$ only increases.
Once again, since our chosen values are optimal, we assume this is already the case.
We substitute for $t_1(u_0) = 0$ to infer that 
\[
3^{t_0}N(u_0)\le 6^{t_0}
\]
where the only constrain is $4t_0 = n$.
Hence,
\[
3^{t_0}N(u_0)\le 6^{t_0} = 6^{n/4}
\]
as desired.

\item[Case 2.]
$m'_R(u_0) / 2 \le t_0 - (m_B(u_0) - t_1(u_0))$.
Let $x = m'_R(u_0)$.
Now, consider increasing $t_1$ by $x/2$ and decreasing $m'_R$ by $x$.
As $\frac{5}{2} \ge \left(\frac{19\sqrt{6}}{30}\right)^2$, such a transformation would only increase the value of $3^{t_0}\cdot N(u_0)$.
This transformation still maintains the invariant that $I(u_0) = n$ and also, $m_B(u_0) + t_1(u_0) \le t_0$.
Since our chosen values are optimal, we assume this is already the case.
Substituting for $m'_R = 0$, we get that 
\[
3^{t_0} N(u_0)
\le \left(\frac{3\sqrt{6}}{2}\right)^{t_0}\left(\frac{5}{2}\right)^{t_1(u_0)} \left(\frac{2\sqrt{6}}{3}\right)^{m_B(u_0)}\\
\]
with the inequalities in the remaining variables being: $3t_0 + 2t_1(u_0) + m_B(u_0) = n, m_B(u_0) + t_1(u_0) \le t_0$

Now, let $x = t_0 - (m_B(u_0) + t_1(u_0))$.
Consider the transformation where we increase $m_B(u_0)$ by $2x$ and decrease $t_1(u_0)$ by $x$.
As $\left(\frac{2\sqrt{6}}{3}\right)^2 \ge \frac{5}{2}$, this transformation can only increase the value of $3^{t_0} N(u_0)$.
This transformation still maintains the invariant that $I(u_0) = n$ and at the end, we have that $m_B(u_0) + t_1(u_0) = t_0$.
Since our chosen values are optimal, we assume this is already the case.
We substitute for $m_B(u_0)$ to get that 
\[
3^{t_0} N(u_0)
\le 6^{t_0}\left(\frac{5\sqrt{6}}{8}\right)^{t_1(u_0)}
\]
where the only constrain is $4t_0 + t_1(u_0) = n$.
As $6 > \left(\frac{5\sqrt 6}{8}\right)^4$, if $t_1(u_0) = x$, then the transformation where we decrease $t_1(u_0)$ by $x$ and increase $t_0$ by $x/4$ maintains the invariant $4t_0 + t_1(u_0) = n$ and the value of $3^{t_0}N(u_0)$ only increases.
Once again, since our chosen values are optimal, we assume this is already the case.
We substitute for $t_1(u_0) = 0$ to get that 
\[
3^{t_0}N(u_0)\le 6^{t_0}
\]
where the only constrain is $4t_0 = n$.
Hence,
\[
3^{t_0}N(u_0)\le 6^{t_0} = 6^{n/4}
\]
as desired.
\end{enumerate}
\end{proof}

\subsubsection{Bounding survival probability in arbitrary stage}\label{subsec: appendix proof bound in stage b}

We prove the bound on conditional survival probability in arbitrary stage:

\begin{proof}[Proof of \cref{lemma: bounding Mwdh by Fwdh}]
By considering cases on the number of markings at the root of a transversal tree, we see that $M(w, d, h)$ obeys the following recurrence:
\begin{align*}
M(w, d, h) =     
\max\{
& (9/4) M(w-1, d-1, h)\\
& 2 M(w-2, d-1, h-1)\\
& (7/4) M(w-2, d-1, h)\\
& (3/2) M(w-3, d-1, h)\\
\}\\
\end{align*}

with the base cases defined as follows:

\[
M(w, 0, h) =
\begin{cases}
1 & w \le 0, h \ge 0\\
0 & \textrm{otherwise}
\end{cases}
\]

For $1\le i\le 4$, define functions $G_i(w, d, h)$ as follows:
\begin{itemize}
\item
$G_1(w, d, h) = \left(\frac{9}{4}\right)^d$.

\item
$G_2(w, d, h) = \left(\frac{9}{4}\right)^{2d-w}(2)^{w-d}$.

\item
$G_3(w, d, h) = \left(\frac{9}{4}\right)^{2d-w}(2)^{h}\left(\frac{27}{8}\right)^{(w-d-h)/2}
= (2)^{h}\left(\frac{27}{8}\right)^{(3d-w-h)/2}\left(\frac{3}{2}\right)^{w-2d}$.

\item
$G_4(w, d, h) = (2)^{3d-w}\left(\frac{3}{2}\right)^{w-2d}$
\end{itemize}

Let $G(w, d, h) = \min_{1\le i\le 4}(G_i(w, d, h))$.

The result follows from the following claims:

\begin{claim}\label{claim: t0 < n/4: G = F}
$G(w, d, h) = F(w, d, h)$.
\end{claim}

\begin{claim}\label{claim: t0 < n/4: M <= Gi}
For $1\le i\le 4$, $M(w, d, h) \le G_i(w, d, h)$.
\end{claim}

Indeed once we prove these claims, we can conclude that for all values of $w, d, h$ and for all $i$, $M(w, d, h)\le G_i(w, d, h)$, implying that for all values of $w, d, h$, $M(w, d, h)\le G(w, d, h) = F(w, d, h)$ as desired.

We now prove these claims one by one:
\begin{proof}[Proof of \cref{claim: t0 < n/4: G = F}]

To prove this claim, we will show the following claims:
\begin{claim}\label{claim: t0 < n/4: G = F: G1 <= G2}
$G_1(w, d, h)\le G_2(w, d, h)$ iff $w\le d$
\end{claim}

\begin{claim}\label{claim: t0 < n/4: G = F: G2 <= G3}
$G_2(w, d, h)\le G_3(w, d, h)$ iff $w\le d+h$
\end{claim}

\begin{claim}\label{claim: t0 < n/4: G = F: G3 <= G4}
$G_3(w, d, h)\le G_4(w, d, h)$ iff $w\le 3d-h$
\end{claim}

These claims together imply that $G = G_1$ if $w\le d$, $G = G_2$ if $d\le w\le d+h$, $G = G_3$ if $d+h\le w\le 3d-h$, and $G = G_4$ if $3d-h\le w$. This exactly matches the definition of $F$ and hence, $G = F$ as desired.

We now compute and prove these claims one by one:

\begin{proof}[Proof of \cref{claim: t0 < n/4: G = F: G1 <= G2}]
We compute that 
\[
\frac{G_1(w, d, h)}{G_2(w, d, h)}
= \left(\frac{9}{8}\right)^{w-d}
\]
Hence, $\frac{G_1(w, d, h)}{G_2(w, d, h)}\le 1$ if and only if $w\le d$ as desired.
\end{proof}

\begin{proof}[Proof of \cref{claim: t0 < n/4: G = F: G2 <= G3}]
We compute that 
\[
\frac{G_2(w, d, h)}{G_3(w, d, h)}
= 2^{w-d-h} \left(\frac{8}{27}\right)^{(w-d-h)/2}
= \left(\frac{32}{27}\right)^{(w-d-h)/2}
\]
Hence, $\frac{G_2(w, d, h)}{G_3(w, d, h)}\le 1$ if and only if $w\le d+h$ as desired.
\end{proof}

\begin{proof}[Proof of \cref{claim: t0 < n/4: G = F: G3 <= G4}]
We compute that 
\[
\frac{G_3(w, d, h)}{G_4(w, d, h)}
= \left(\frac{27}{32}\right)^{(3d-w-h)/2}
\]
Hence, $\frac{G_3(w, d, h)}{G_4(w, d, h)}\le 1$ if and only if $w\le 3d-h$ as desired.
\end{proof}

\end{proof}

\begin{proof}[Proof of \cref{claim: t0 < n/4: M <= Gi}]

To prove this claim, we will show the following claims:
\begin{claim}\label{claim: t0 < n/4: G = F: M <= G1}
For all $w, d, h$: $M(w, d, h)\le G_1(w, d, h)$.
\end{claim}

\begin{claim}\label{claim: t0 < n/4: G = F: M <= G2}
For all $w, d, h$: $M(w, d, h)\le G_2(w, d, h)$.
\end{claim}

\begin{claim}\label{claim: t0 < n/4: G = F: M <= G3}
For all $w, d, h$: $M(w, d, h)\le G_3(w, d, h)$.
\end{claim}

\begin{claim}\label{claim: t0 < n/4: G = F: M <= G4}
For all $w, d, h$: $M(w, d, h)\le G_4(w, d, h)$.
\end{claim}

Using these claims, the over claim indeed follows.

We now prove these claims one by one:

\begin{proof}[Proof of \cref{claim: t0 < n/4: G = F: M <= G1}]
We proceed by induction on $d$. For $d = 0$, the claim trivially holds.    
For $d\ge 1$, we compute:
\begin{align*}
M(w, d, h)
& = \max((9/4)M(w-1, d-1, h), 2M(w-2, d-1, h-1),\\
& (7/4)M(w-2, d-1, h), (3/2)M(w-3, d-1, h))\\
& = \max((9/4)G_1(w-1, d-1, h), 2G_1(w-2, d-1, h-1),\\
& (7/4)G_1(w-2, d-1, h), (3/2)G_1(w-3, d-1, h))\\
& \le \max\left((9/4)\left(\frac{9}{4}\right)^{d-1}, 2\left(\frac{9}{4}\right)^{d-1}, (7/4)\left(\frac{9}{4}\right)^{d-1}, (3/2)\left(\frac{9}{4}\right)^{d-1}\right)\\
& = \left(\frac{9}{4}\right)^{d}\max\left(1, \frac{8}{9}, \frac{7}{9}, \frac{2}{3}\right)\\
& = \left(\frac{9}{4}\right)^{d}\\
& = G_1(w, d, h)
\end{align*}
\end{proof}

\begin{proof}[Proof of \cref{claim: t0 < n/4: G = F: M <= G2}]
We proceed by induction on $d$. For $d = 0$, the claim trivially holds.    
For $d\ge 1$, we compute:
\begin{align*}
M(w, d, h)
& = \max((9/4)M(w-1, d-1, h), 2M(w-2, d-1, h-1),\\
& (7/4)M(w-2, d-1, h), (3/2)M(w-3, d-1, h))\\
& = \max((9/4)G_2(w-1, d-1, h), 2G_2(w-2, d-1, h-1),\\
& (7/4)G_2(w-2, d-1, h), (3/2)G_2(w-3, d-1, h))\\
& \le \max\biggl((9/4)\left(\frac{9}{4}\right)^{2d-w-1}(2)^{w-d}, 2\left(\frac{9}{4}\right)^{2d-w}(2)^{w-d-1},\\
& (7/4)\left(\frac{9}{4}\right)^{2d-w}(2)^{w-d-1}, (3/2)\left(\frac{9}{4}\right)^{2d-w+1}(2)^{w-d-2}\biggr)\\
& \le \left(\frac{9}{4}\right)^{2d-w}(2)^{w-d}\max\left(1, 1, \frac{7}{8}, \frac{27}{32}\right)\\
& \le \left(\frac{9}{4}\right)^{2d-w}(2)^{w-d}\\
& = G_2(w, d, h)
\end{align*}
\end{proof}

\begin{proof}[Proof of \cref{claim: t0 < n/4: G = F: M <= G3}]
We proceed by induction on $d$. For $d = 0$, the claim trivially holds.    
For $d\ge 1$, we compute:
\begin{align*}
M(w, d, h)
& = \max((9/4)M(w-1, d-1, h), 2M(w-2, d-1, h-1),\\
& (7/4)M(w-2, d-1, h), (3/2)M(w-3, d-1, h))\\
& = \max((9/4)G_3(w-1, d-1, h), 2G_3(w-2, d-1, h-1),\\
& (7/4)G_3(w-2, d-1, h), (3/2)G_3(w-3, d-1, h))\\
& \le \max\biggl((9/4)\left(\frac{9}{4}\right)^{2d-w-1}(2)^{h}\left(\frac{27}{8}\right)^{(w-d-h)/2},
2\left(\frac{9}{4}\right)^{2d-w}(2)^{h-1} \left(\frac{27}{8}\right)^{(w-d-h)/2},\\
& (7/4)\left(\frac{9}{4}\right)^{2d-w}(2)^{h}\left(\frac{27}{8}\right)^{(w-d-h-1)/2}, (3/2)(2)^h\left(\frac{27}{8}\right)^{(3d-w-h)/2}\left(\frac{3}{2}\right)^{w-2d-1}\biggr)\\
& \le G_3(w, d, h) \max\left(1, 1, \left(\frac{49}{54}\right)^{1/2}, 1\right)\\
& = G_3(w, d, h)
\end{align*}
\end{proof}

\begin{proof}[Proof of \cref{claim: t0 < n/4: G = F: M <= G4}]
We proceed by induction on $d$. For $d = 0$, the claim trivially holds.    
For $d\ge 1$, we compute:
\begin{align*}
M(w, d, h)
& = \max((9/4)M(w-1, d-1, h), 2M(w-2, d-1, h-1),\\
& (7/4)M(w-2, d-1, h), (3/2)M(w-3, d-1, h))\\
& = \max((9/4)G_4(w-1, d-1, h), 2G_4(w-2, d-1, h-1),\\
& (7/4)G_4(w-2, d-1, h), (3/2)G_4(w-3, d-1, h))\\
& \le \max\biggl((9/4)(2)^{3d-w-2} \left(\frac{3}{2}\right)^{w-2d+1}, 2(2)^{3d-w-1}\left(\frac{3}{2}\right)^{w-2d},\\
& (7/4)(2)^{3d-w-1}\left(\frac{3}{2}\right)^{w-2d}, (3/2)(2)^{3d-w}\left(\frac{3}{2}\right)^{w-2d-1}\biggr)\\
& \le (2)^{3d-w}\left(\frac{3}{2}\right)^{w-2d}\max\left(\frac{27}{32}, 1, \frac{7}{8}, 1\right)\\
& \le (2)^{3d-w}\left(\frac{3}{2}\right)^{w-2d}\\
& = G_4(w, d, h)
\end{align*}
\end{proof}

\end{proof}

\end{proof}

\end{document}